\documentclass[journal,10pt]{IEEEtran}

\pdfoutput=1
\usepackage{times,amsmath,epsfig}

\newif\iffullversion
\fullversiontrue 

\newif\ifdraft
\draftfalse 

\newif\ifsubmission
\submissiontrue 

\newif\ifplot
\plottrue

\newif\ifprotocol
\protocoltrue

\usepackage{url} 
\usepackage{xfloat}
\usepackage{varioref}
\usepackage{ifthen}
\usepackage{color}
\usepackage{graphicx}

\usepackage[english]{babel}
\usepackage[latin1]{inputenc}
\usepackage{epsfig}
\usepackage{amssymb}
\usepackage{amsmath}
\usepackage{amsfonts}
\usepackage{amstext}
\usepackage{amsbsy}
\usepackage{psfrag,pstricks}
\usepackage{verbatim}
\usepackage{moreverb}
\usepackage{stmaryrd}
\usepackage{multirow}
\usepackage{multicol}
\usepackage{tabularx}
\usepackage{dcolumn}
\usepackage{subfigure}
\usepackage[titles,subfigure]{tocloft}
\usepackage{float}
\usepackage{subfloat}

\ifsubmission
  \newcommand{\TODO}[1]{}
  
  \newcommand{\QUESTION}[1]{}
  \newcommand{\CHECK}[1]{}
  
  \newcommand{\ADDED}[1]{}
  \newcommand{\NOTE}[1]{}
  \newcommand{\ANSWER}[1]{}
\else
  \newcommand{\TODO}[1]{\textcolor{red}{{TO DO:#1}}}
  
  \newcommand{\QUESTION}[1]{\\\textcolor{red}{?: {#1}}}
  \newcommand{\CHECK}[1]{\\\textcolor{blue}{Please check: {#1}}}
  
  \newcommand{\ADDED}[1]{\textcolor{blue}{{#1}}}
  \newcommand{\NOTE}[1]{{\it\textcolor{red}{{#1}}}}
  \newcommand{\ANSWER}[1]{{\it\textcolor{purple}{{#1}}}}
\fi

\newcommand{\Z}{\mathbb Z}

\newcommand{\enc}[1]{\llbracket #1 \rrbracket}

\newcommand{\sect}[1]{Section~\ref{#1}}

\newcommand{\fig}[1]{Figure~\ref{#1}}
\newcommand{\tab}[1]{Table~\ref{#1}}
\newcommand{\xor}{\oplus}

\newcommand{\domain}{\mathtt{Dom}} 

\makeatletter
\newcommand{\thickhline}{%
    \noalign {\ifnum 0=`}\fi \hrule height 1pt
    \futurelet \reserved@a \@xhline
}
\newcolumntype{"}{@{\hskip\tabcolsep\vrule width 1pt\hskip\tabcolsep}}
\makeatother

\newlistof{pprotocol}{lop}{List of Protocols}

\floatstyle{boxed}
\newfloat{ppp}{htb}{}
\floatname{ppp}{Protocol}

\title{Piecewise Function Approximation with \\ Private Data}
\author{%
{Riccardo Lazzeretti{\small $~^{1}$}\thanks{\textcolor{red}{\it A revised version of the paper will be soon submitted to IEEE Transaction on Information Forensic and Security.}}, Tommaso Pignata{\small $~^{2}$}, Mauro Barni{\small $~^{3}$}}%
\vspace{1.6mm}\\
\fontsize{10}{10}\selectfont\itshape
Department of Information Engineering and Mathematics, University of Siena, Italy\\
\fontsize{9}{9}\selectfont\ttfamily\upshape
$^{1}$\,lazzeretti@diism.unisi.it,
$^{2}$\,pignata.tommaso@gmail.com,
$^{3}$\,barni@dii.unisi.it%
}

\begin{document}
%
\maketitle

\begin{abstract}
We present two Secure Two Party Computation (STPC) protocols for piecewise function approximation on private data. The protocols rely on a piecewise approximation of the to-be-computed function easing the implementation in a STPC setting. The first protocol relies entirely on Garbled Circuit (GC) theory, while the second one exploits a hybrid construction where GC and Homomorphic Encryption (HE) are used together. In addition to piecewise constant and linear approximation, polynomial interpolation is also considered. From a communication complexity perspective, the full-GC implementation is preferable when the input and output variables can be represented with a small number of bits, while the hybrid solution is preferable otherwise. With regard to computational complexity, the full-GC solution is generally more convenient.
\end{abstract}


 \section{Introduction}\label{sec:intro}
    \IEEEPARstart{T}{he} interest towards applications where two or more non-trusted parties wish to collectively process one or more signals to reach a common goal has prompted the quest of tools and protocols capable of processing signals and data directly in the encrypted domain \cite{erkin2007protection,SPM13}. The availability of such protocols would avoid that non-trusted parties refuse to cooperate even when they have a common goal because they are not willing to disclose their private inputs to the other parties. Processing signals directly in the encrypted domain, in fact, allows each party to observe only its own input and its share of the computation output, thus avoiding the need to disclose sensitive information to the others. In recent literature, protocols like those described above are often referred to as Signal Processing in the Encrypted Domain (s.p.e.d.). The number of possible applications of s.p.e.d. is virtually endless. Among the most interesting scenarios investigated so far we mention: private data mining \cite{agrawal2000privacy}, secure processing of biometric data \cite{bringer2008authentication,erkin2009privacy,luo2012efficient}, secure processing of biomedical signals \cite{barni2011privacyECG,lazzeretti2012privacy}, processing of private user preferences \cite{beye2011efficient}, fusion of private data \cite{lazzeretti2014gossip}, etc.

From a technical point of view, s.p.e.d. protocols rely on Secure Multi-Party Computation (SMPC), a cryptographic discipline rooted in the seminal works by Goldreich \cite{goldreich1987play}, Rivest et al. \cite{rivest1978method} and Yao \cite{Yao82}. In the simplest case, like the one considered in this paper, the protocol involves only two parties. In this case, we speak about Secure Two-Party Computation (STPC) instead of SMPC. In a general STPC setting, one party, say Alice, owns a signal that must be processed in some way by the other party, hereafter referred to as Bob. Bob must process Alice's signal without getting any information about it, in some scenarios not even the result of the computation. At the same time, Bob is interested to protect the information he uses to process the signal. 

Several cryptographic primitives for STPC exist, which when coupled with a suitable design of the underlying signal processing algorithms, permit to process the signals in a secure way. 
The two main approaches to STPC are based, respectively, on Homomorphic Encryption (HE) and Garbled Circuits (GCs).
HE provides a simple and elegant way to evaluate linear operations on encrypted data \cite{rivest1978method}, however when non linear operations are involved, it is necessary to resort to ad-hoc, interactive and usually complex protocols. For instance, researchers have developed HE-based protocols to compute common non linear functions such as bit decomposition and comparison \cite{erkin2009privacy}, division \cite{veugen2010encrypted},  etc. On the other hand, GCs allow to evaluate any function that can be represented with an acyclic boolean circuit. For this reason GCs are very powerful tools when the functionalities involved in the computation can be represented by simple circuits, like in the case of comparison, addition, multiplexing, etc.. In other cases, however, the boolean circuit required to describe the functionality is so complex to make the use of GCs problematic. This is the case, for instance of product, division \cite{lazzeretti2011division}, logarithm \cite{barni2010privacy} computation, etc. Given the complementary pros and cons of HE and GC, the use of hybrid protocols has been proposed to take advantage of the benefits offered by the two approaches, so that complex protocols are developed as a concatenation of sub-protocols, some of them implemented by using HE and others by using GCs \cite{barni2011privacyECG,lazzeretti2012privacy}. A simple protocol to securely link HE and GC subprotocols is described in \cite{kolesnikovdust}.

Recently, Fully Homomorphic Encryption (FHE) schemes \cite{gentry2009fully,van2010fully,gentry2011implementing} have been proposed, allowing the evaluation of both addition and product between encrypted values on the service provider side, without any interaction with the party owning the public key, whose only tasks is to encrypt the inputs and decrypt the results. Unfortunately, FHE is still highly inefficient, principally due to the huge size of the public key. 


Regardless of the adopted approach, computing a generic function, like trigonometric, hyperbolic and statistical functions on private data is a difficult problem. Classical HE schemes (like \cite{Pail99,elg85}) do not provide a general approach for universal function evaluation, while both GC and FHE require that the to-be-computed function is described as a logical circuit working on binary variables, thus making the implementation of complex functions like trigonometric functions, extremely inefficient.

With the above ideas in mind, the main contribution of this paper is the proposal of a new class of protocols that permit to evaluate privately a piecewise constant, linear or polynomial approximation of any function $f()$ having limited domain and codomain, for any given choice of the approximation parameters like the representation error and the accuracy of the input and output variables. To start with, the piecewise approximation is determined, specifying the sub-intervals the function domain is split into to define the piecewise approximation, and the approximation parameters to be used within each subinterval. Then, given the input, the actual function approximation consists of three main steps: interval detection, parameter retrieval and approximation. In the first step the domain interval the input belongs to is detected, then the parameters of the (constant, linear o polynomial) approximation are retrieved and finally the actual approximating value is evaluated.


The present work generalizes and improves the system proposed in \cite{pignata2012general}, where piecewise linear function approximation by means of full-GC or Hybrid protocols is considered. As a first main difference, in this paper we consider a general approximation framework, which is not limited to piecewise linear functions, but includes also polynomial approximation. Secondly, and equally important, the efficiency of the protocols proposed in \cite{pignata2012general} is significantly improved with regard to both the full-GC and hybrid implementations. Specifically, the complexity of the segment detection phase is decreased from $\mathcal{O}(N\ell)$ to $\mathcal{O}(N)$, where $N$ is the number of segments and $\ell$ the input bitlength, and the complexity of the parameter retrieval phase is diminished down to a point to become negligible.

With respect to the two approaches proposed in the paper, one fully based on GC and the other relying on a hybrid protocol, we show that from a communication point of view the hybrid solution is preferable when the input is represented with a large number of bits and a high precision is needed, otherwise the full-GC solution is preferable. With regard to computational complexity, our implementations reveal that full-GC solution is always preferable.

This paper is organized as follows. In \sect{sec:tools}, the main cryptographic tools the protocols rely on are presented. The general framework for piecewise polynomial approximation is presented in \sect{sec:approximation}, together with the instantiations of such a framework for specific cases of particular interest. 
The number of bits that must be used to represent the protocol variables is evaluated in \sect{sec:error} as a function of the approximation error.
Different STPC implementations for various classes of approximation functions are presented in \sect{sec:protocols} and compared in \sect{sec:comparing}. The paper ends with some conclusions in \sect{sec:conclusion}.

 \section{Cryptographic Tools}\label{sec:tools}
   In this section, we present the cryptographic primitives at the bases of our protocols, namely Oblivious Transfer (OT), Garbled Circuits (GC) and Homomorphic Encryption (HE). 

Throughout the paper we adopt the semi-honest security model, where the parties involved are assumed to follow the protocol as prescribed but try to learn as much as possible from the exchanged messages and their private inputs.
While the security of the single tools in the semi-honest setting are demonstrated in the original papers, the security of their composition in hybrid protocols is proven in \cite{kolesnikovdust}.

\subsection{Homomorphic Encryption}
With a semantically secure, additively homomorphic, asymmetric encryption scheme, it is possible to obtain the encryption of the sum of two values $a$ and $b$ available in encrypted form\ through the product of the corresponding ciphertexts. In other words, by denoting with $\enc{\cdot}$ the encryption operator, we have $\enc{a+b}=\enc{a}\cdot\enc{b}$. In a similar way it is possible to compute the product between two values, one of them available in non-encrypted form, through exponentiation, i.e. $\enc{ab}=\enc{a}^b$. More complex functionalities, such as bit decomposition \cite{schoenmakers2006efficient} and comparison \cite{damgard2007efficient}, can also be evaluated by interacting with the owner of the decryption key in protocols characterized by a rather high complexity in terms of protocol rounds and number of transmitted cyphertexts.

The most widely used additively homomorphic cryptosystem is Paillier cryptosystem \cite{Pail99} with plaintext space $\Z_N$ and ciphertext space $\Z_{N^2}^*$, where $N$ is a $T$-bit RSA modulus and a ciphertext is represented with $2T$ bits. The communication complexity of HE-based protocols  is mainly related to the number of cyphertexts to be transmitted and the number of rounds necessary for the protocol evaluation. The computational complexity is usually measured in terms of number of modular exponentiations, and by considering that encryption and decryption have a complexity similar to exponentiations.

\subsection{Oblivious Transfer}
Oblivious Transfer (OT) protocols \cite{even1985randomized} allow one party, the chooser, to select one out of two (or more) inputs provided by another party, the sender, in a way that protects both parties: the sender is assured that the chooser does not receive more information than it is entitled, while the chooser is assured that the sender does not learn which input he received. 

OT protocols can be subdivided in two phases: the off-line and online phases. It is customary to move the set up operations and a great part of the most computationally expensive operations to the offline phase, which is performed during inactivity, and during which the chooser and the sender evaluate many OTs computed on random values. Then, during the online phase the result of precomputed OTs are used to update the OTs to the actual values \cite{Beaver95}.
During the offline phase a great number of OTs can be evaluated in parallel in $3$ communication rounds, transmitting $~6t$ bits for each OT, where $t$ is the input bitlength, while the online phase needs only the transmission of $~2t$ bits in $2$ rounds for each OT \cite{IKNP03}.

\subsection{Garbled Circuits}
\begin{figure}[!hbt]
\centering
\ifdraft
\includegraphics[width=0.45\columnwidth]{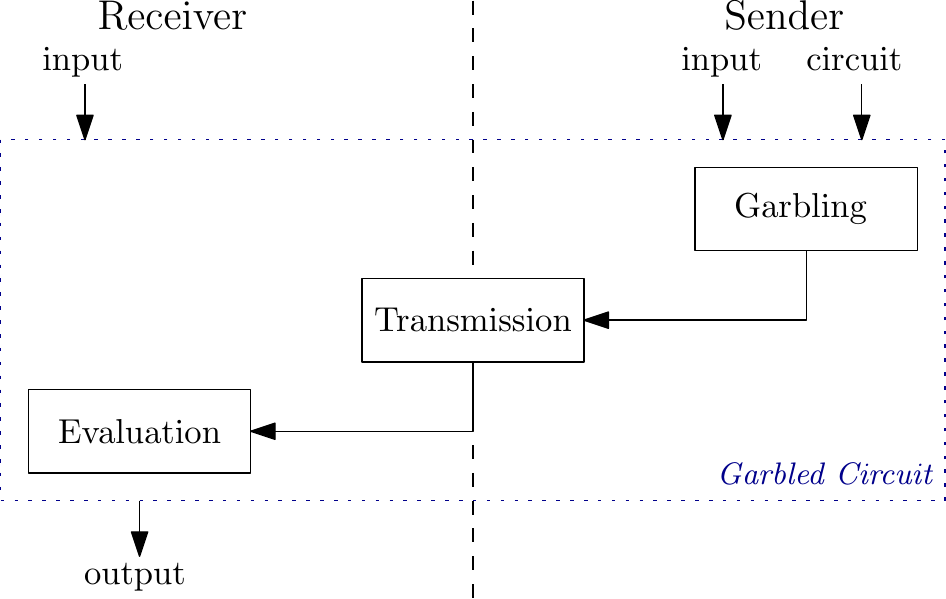}
\else
\includegraphics[width=0.81\columnwidth]{GC}
\fi
\caption{Garbled circuit scheme.}
\label{fig:gc}
\end{figure}

Any boolean circuit containing no cycle can be privately evaluated on secret inputs by using Garbled Circuits (GC). 
Despite several optimizations proposed later, the overall protocol for GC evaluation is still similar to the first one proposed by Yao \cite{Yao82,Yao86How}.
As shown in \fig{fig:gc}, a GC is evaluated in three steps. 
During garbling, one party,  the sender, associates a couple of secrets (one for each logical value) to each wire of the circuit and garbles each gate by encrypting, for each row of the corresponding truth table, the secret associated to the output by using the two secrets associated to the inputs.
In the transmission phase, the sender transmits the garbled tables to the other party, the receiver. Moreover the sender transmits the secrets associated to the input wires linked to his inputs, while the receiver obtains the secrets of his input wires by performing an OT together with the sender.
Finally, during evaluation, the receiver decrypts the secrets gate by gate starting from the gates connected to the inputs (for a detailed description of each of the above steps we refer to \cite{lazzeretti2013private}).

Thanks to recent optimizations \cite{MNPS04,KS08XOR,pinkas2009practical}, garbling, transmission and evaluation of XOR gates have negligible complexity, while for each non-XOR binary gate circuit garbling requires the computation of $3$ Hash functions and the transmission of $3t$ bits, where $t$ is a security parameter (usually $t=80$ for short term security). In addition, gate evaluation requires the computation of a Hash function with probability $3/4$. For each input bit of the sender ,a secret of $t$ bits is transmitted, while for each input bit of the receiver an OT is evaluated ($2t$ bits transmitted online).
We underline that if the sender and the receiver know in advance the functionality to be evaluated, garbling and circuit transmission can be performed offline, when the inputs are not yet available.

\subsection{Hybrid protocols}\label{sec:hybrid}
The use of hybrid protocols, such as in \cite{barni2011privacyECG,barni2010privacy}, permits to efficiently evaluate functionalities for which full-HE or full-GC solutions would not be efficient (or even impossible). Given that GC and HE rely on different ways of representing data, conversion from homomorphic ciphertexts to garbled secrets (or vice versa) must be performed by resorting to interfacing protocols based on additive blinding. In particular, by referring to the protocols described in \cite{kolesnikovdust}, it is easy to derive that the conversion of an $\ell$-bit long value from HE to GC requires the on-line transmission of additional $2T+7\ell t$ bits, while conversion from GC to HE requires an overhead of $2T+(\ell+\tau) 5t$ bits, where $\tau$ is an obfuscation security parameter (usually $\tau=80$).

 \section{Function approximation}\label{sec:approximation}
	
\begin{figure}[hbt!]
\centering
\ifdraft
\includegraphics[width=.5\columnwidth]{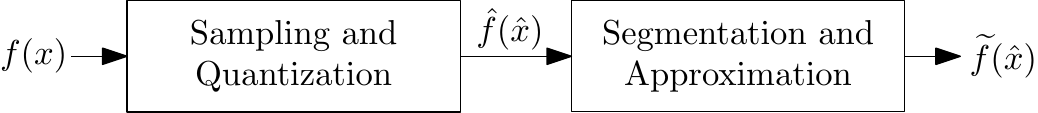}
\else
\includegraphics[width=\columnwidth]{toApprox}
\fi
\caption{High level steps to map a function $f(x)$ into a piecewise polynomial function $\widetilde{f}(\hat{x})$ in a discrete space.}
\label{fig:toApprox}
\end{figure}

Given a generic limited function $f()$ with domain $\domain=[x_a,x_b)$ and codomain $[y_a,y_b)$, our goal is to find a way to approximate $f()$ in $\domain$ so that the approximation can be efficiently evaluated in the encrypted domain, as shown later in \sect{sec:protocols}.
The solutions we focus on in this paper permit to represent a sampled and quantized version of $f()$  - say  $\hat{f}()$ - through a piecewise polynomial function $\widetilde{f}()$, as shown in \fig{fig:toApprox}. To be specific, the approximation procedure we propose consists of two main steps. First the domain of $\hat{f}()$ is partitioned into a given number of non-overlapping intervals. Then, for each interval, a polynomial is chosen to approximate $\hat{f}()$. 

\subsection{Quantization}
First of all, considering that STPC protocols work with integer values, we introduce a discretized version of $f()$. To do so, we assume that the input variable $x$ is represented by $\ell_x$ bits and the function output $y$ with $\ell_y$ bits. We also find it convenient to translate and scale the domain and codomain of $f()$ and to represent input and output by using integer numbers so that the input ranges in the interval $[0,n)\cap\mathbb{N}$ and the output in the interval $[0,m)\cap\mathbb{N}$, where $n=2^{\ell_x}$ and $m=2^{\ell_y}$. More specifically, we define the normalized and quantized input and output variables respectively as $\hat{x} = \left\lfloor q_x (x-x_a)\right\rfloor$, with $q_x = \frac{2^{\ell_x}}{x_b-x_a}$ and $\hat{y} = \left\lfloor q_y (y-y_a)\right\rfloor$ with $q_y = \frac{2^{\ell_y}}{y_b-y_a}$. 
Of course, these operations introduce an approximation error $\epsilon_\ell$ that can be reduced by increasing $\ell_x$ and $\ell_y$. In particular $\ell_y$ must be chosen large enough so that the step used to quantize the output is lower than the desired approximation error, i.e. $1/q_y < \epsilon$. With the above understanding, the normalized and discretized version of $f()$ can be represented by a sequence of value pairs $(\hat{x}_i,\hat{y}_i)$, with $\hat{x}_i = i$, $\forall i = 0 \dots n-1$ and $\hat{y}_i=\left\lceil f\left(\frac{\hat{x}_i}{q_x}+x_a\right) \right\rfloor=\left\lceil f(x)\right\rfloor$. In other words, for any $i$ ranging from $0$ to $n-1$ we have $\hat{y}_i = \hat{f}(\hat{x}_i)$.

\subsection{Domain Partitioning and Polynomial Approximation}

Generally speaking, given the degree of the polynomials used for the approximation, determining the best way of partitioning the domain of $\hat{f}()$ is not an easy task. For this reason, we avoid looking for the optimal segmentation of $\domain$ and restrict our analysis to the identification of a partition that permits to keep the approximation error below a predefined value, while allowing an efficient implementation by relying on STPC techniques. In other words, regardless of the polynomial degree, given an error $\epsilon>1/q_y$,  the goal is to partition the domain into intervals $S_j$, each delimited by the left and right extremes $\hat{s}_j^l$ and $\hat{s}_j^r$, wherein the distance between $\hat{f}(\hat{x}_i)$  and the looked-for approximation $\widetilde{f}(\hat{x}_i)$ is lower than the given error, i.e., considering the amplification factor in the output, $|\tilde{f}(\hat{x}_i)-\hat{f}(\hat{x}_i)|\leq\epsilon q_y$.

To determine a partition of $\domain$ with the desired characteristics, we use a bisection algorithm, inspired by a {\it divide and conquer} strategy. The partitioning algorithm starts with the analysis of the whole segment $S\doteq \domain$, which is subdivided by using the following recursive procedure:

\begin{enumerate}
\item A polynomial approximation of $\hat{f}()$ in $S$ is computed and the maximum error $\epsilon_S=\max_{\hat{x}_i\in S} |\tilde{f}(\hat{x}_i)-\hat{f}(\hat{x}_i)|$ is evaluated;
\item if $\epsilon_S\leq\epsilon q_y$ the parameters of the interpolating polynomial are stored and the set is marked as a leaf of the binary-tree associated to the bisection algorithm, otherwise
\begin{enumerate}
\item  $S$ is marked as a node of the binary-tree and subdivided into 2 subsets each one having size one half of $S$,
\item the procedure is applied from step 1 to each subset.
\end{enumerate}
\end{enumerate}
Note that if a good approximation is not found, the bisection continues until the leaves of the tree coincide with the single points of the discrete domain.
In addition to its simplicity, the above bisection algorithm ensures that the size of all the intervals of the partition is a power of 2, thus enabling an efficient GC implementation of the approximation algorithm, as shown in \sect{sec:GCprimo}.

\begin{figure*}[!bt]
\begin{center}
\subfigure[Original]{\includegraphics[width=0.26\linewidth]{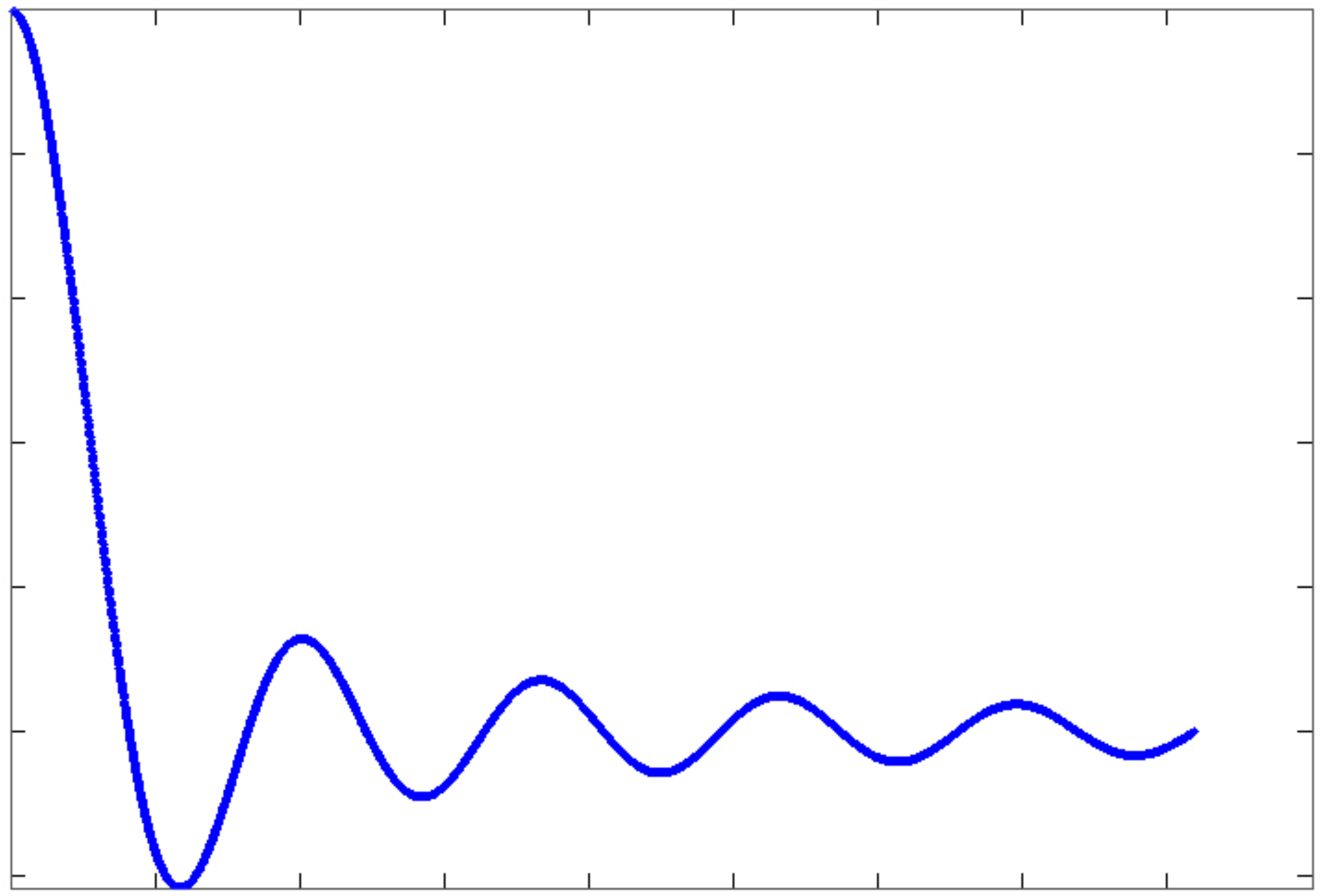}} \hspace*{0.05\linewidth}
\subfigure[Continuous Linear Approximation]{\includegraphics[width=0.26\linewidth]{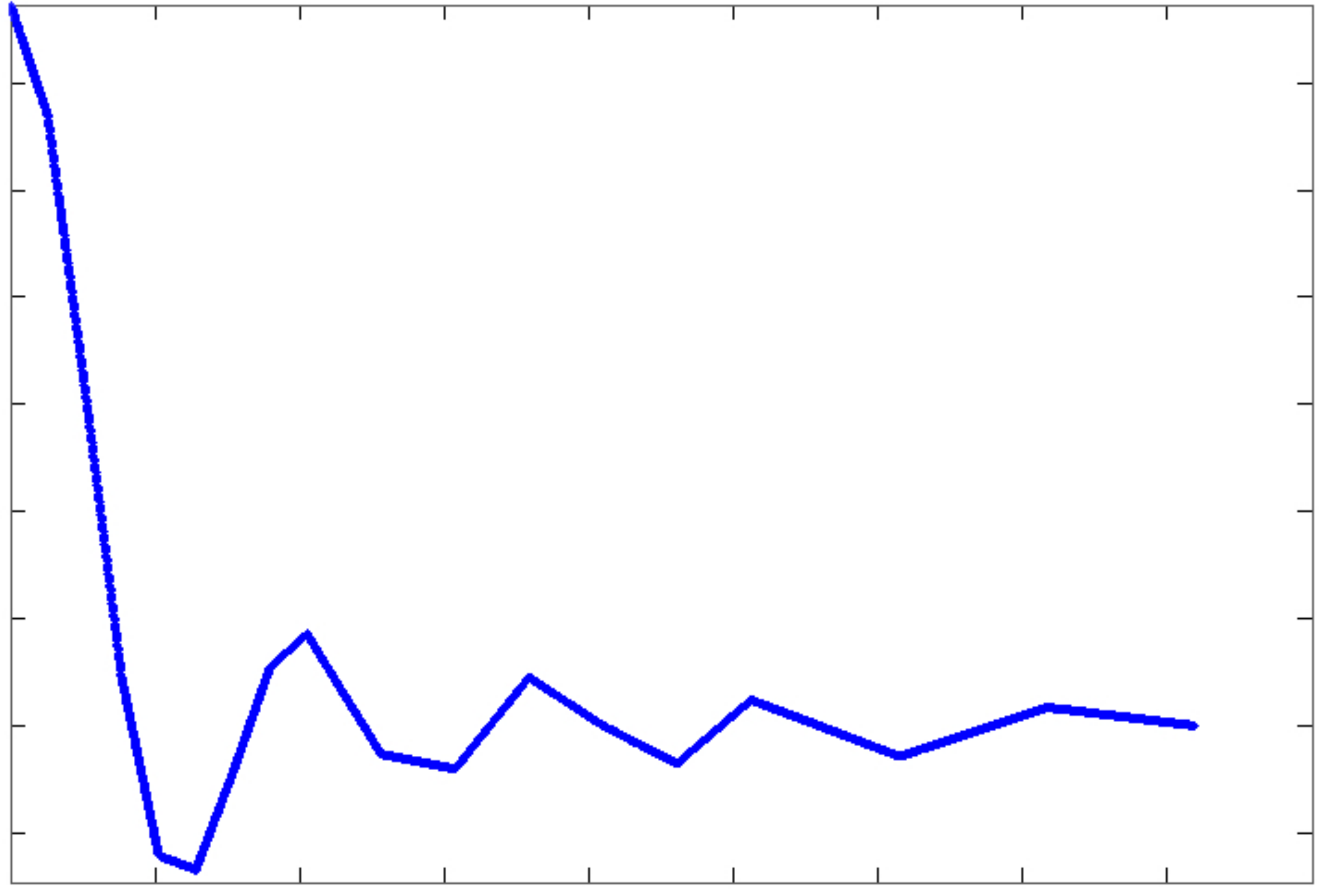}} \hspace*{0.05\linewidth}
\subfigure[Continuous Quadratic Approximation]{\includegraphics[width=0.26\linewidth]{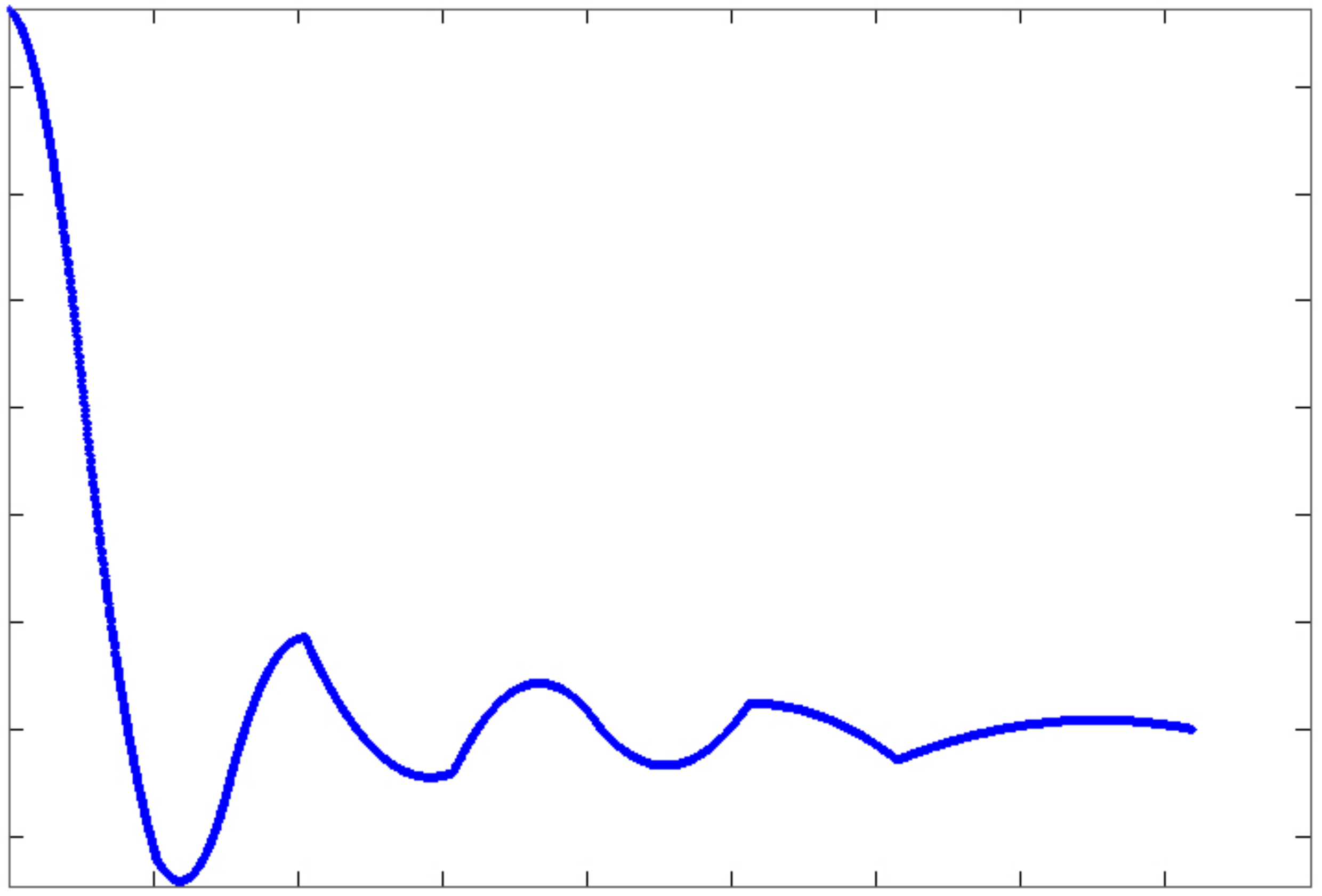}} \\

\subfigure[Constant Approximation]{\includegraphics[width=0.26\linewidth]{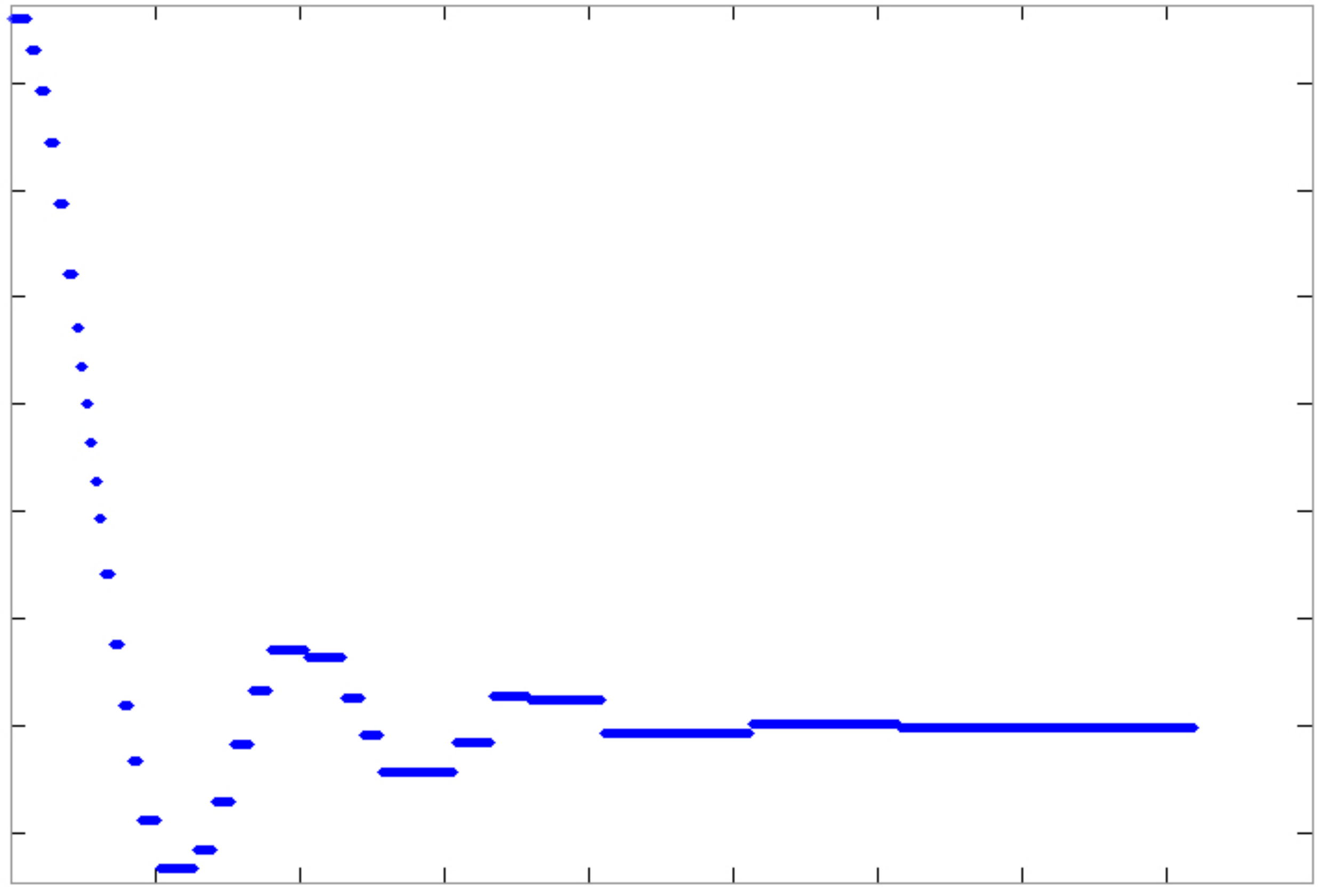}} \hspace*{0.05\linewidth}
\subfigure[Linear Approximation]{\includegraphics[width=0.26\linewidth]{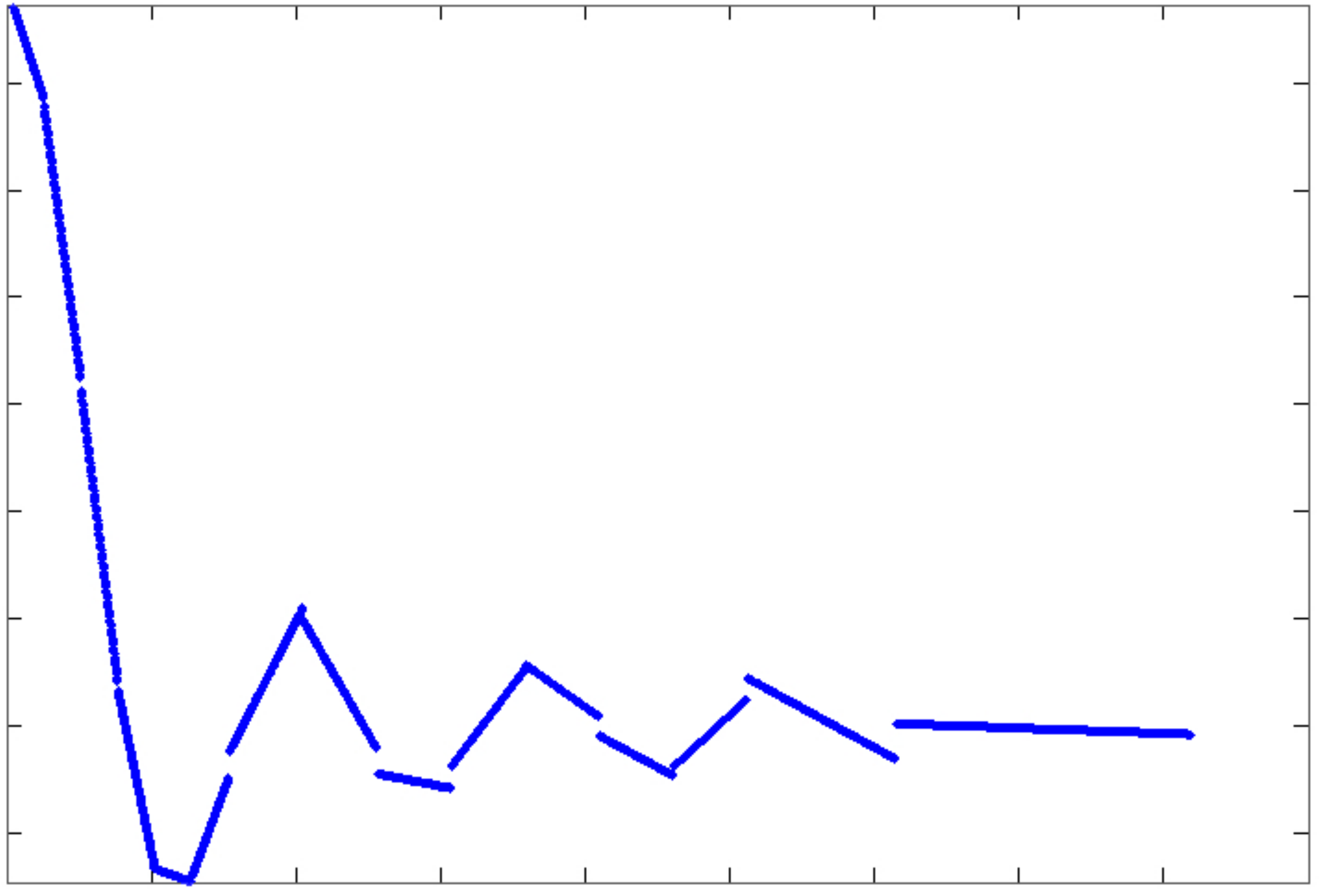}} \hspace*{0.05\linewidth}
\subfigure[Quadratic Approximation]{\includegraphics[width=0.26\linewidth]{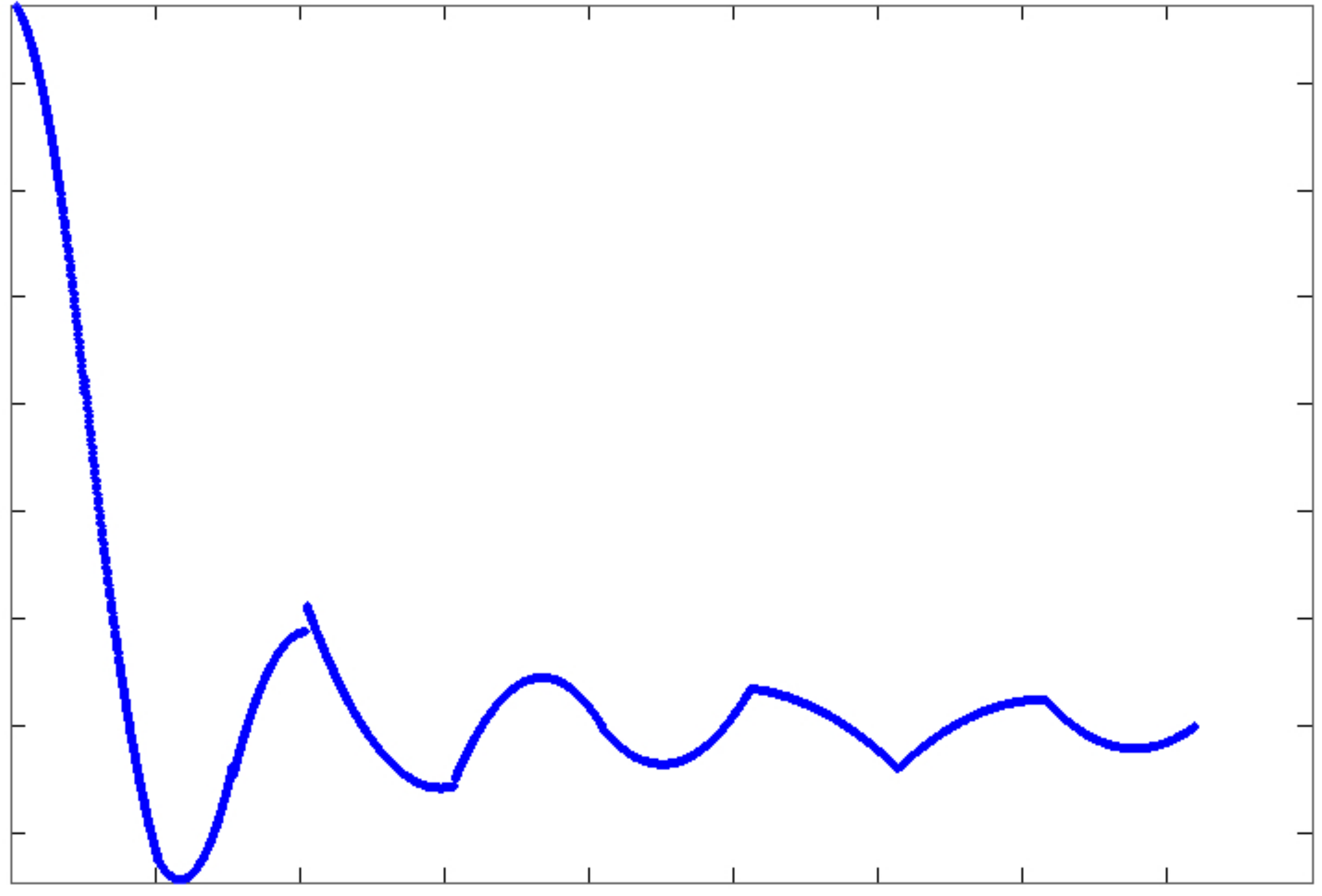}} 

\caption{Approximation of $sinc(x)$ in $[0,10)$ with different classes of interpolating functions.}
\label{fig:approximation01}
\end{center}
\end{figure*}

The depth of the tree and the number of segments of the final partition depends on the approximation type. Here we consider constant, linear and polynomial approximations and exemplify the results by considering the approximation of $sinc(x)$ (see \fig{fig:approximation01}). We assume that approximation parameters are real-valued, postponing the discussion about their integer representation within secure protocols to \sect{sec:error}.

	\subsubsection{Constant approximation}\mbox{} \\*
	In this case $\hat{f}()$ is approximated by a constant. Given a segment $S_j$, the maximum approximation error is minimized by choosing the approximating constant as $$c_j=\frac{\max_{\hat{x}\in S_j} \hat{f}(\hat{x})+\min_{\hat{x}\in S_j} \hat{f}(\hat{x})}{2}.$$

If $|c_j-\hat{f}(\hat{x})|\leq \epsilon q_y$ for all the values $\hat{x}\in S_j$, the approximation is satisfactory, otherwise the segment is subdivided again.
The main advantage of this approximation is its simplicity. A single value is assigned to each leaf of the quad-tree and, once the correct leaf has been selected, the approximated value is immediately obtained, with no additional operations.

	\subsubsection{Linear approximation}\mbox{} \\*
		In each segment $S_j$, $\hat{f}()$ is approximated through a linear function:
$$m_j(\hat{x}-\hat{s}^l_j)+q_j,$$ where $\hat{s}^l_j$ is the left extreme of the segment and $m_j$ and $q_j$ are chosen so to minimize 
$\max_{\hat{x}\in S_j}\{|\hat{f}(\hat{x})-m_j(\hat{x}-\hat{s}^l_j)-q_j|\}$. 
Due to the difficulty of solving the above minimization, we replace it with the search for the regression line that minimizes the square error, i.e. we solve $\min\sum_{\hat{x}\in S_j}(\hat{f}(\hat{x})-m_j(\hat{x}-\hat{s}^l_j)-q_j)^2$.
Then the maximum error in the interval is evaluated and if it is not lower than the desired error the segment is subdivided again.
	\subsubsection{Continuous linear approximation}\label{sec:cla}\mbox{} \\*
	   Assuming that the domain is subdivided into $m$ segments, we may desire to enforce the continuity between the linear approximations used in consecutive segments, even if this can result in a larger number of segments. To do so we need to solve the following linear optimization problem:
\begin{equation}
\left\{\begin{array}{l}
\min\max_{\hat{x}\in Dom}\{|\hat{f}(\hat{x})-m_{j(\hat{x})}(\hat{x}-\hat{s}^l_{j(\hat{x})})-q_{j(\hat{x})}|\}\\ 
q_{j+1}=m_{j}(\hat{s}^r_{j}-\hat{s}^l_{j})+q_{j}\qquad\forall j=0\ldots m-2
\end{array}\right.
\end{equation}
where $j(\hat{x})$ denotes the segment $S_j$ containing $\hat{x}$, while $\hat{s}_j^l$ and $\hat{s}_j^r$ are the left and right extremes of the segment $S_j$.

We can easily observe that the above optimization problem has $2m$ variables and $m-1$ constraints, in addition, if a certain partition does not guarantee an error lower than the target maximum error, some segments must be split and the whole optimization problem has to be solved again. To simplify the problem, we decided to impose that the approximating function is equal to the original one at the extremes of each segment. This is obviously a sub-optimal solution, that, however, permits to compute very quickly the desired approximation. Under this assumption, the continuous linear piecewise approximation is fully defined by the extreme points of the $m$ segments, i.e. 
the $m+1$ couples $(\hat{s}_0^l,\hat{y}_0^l)$ $(\hat{s}_1^l,\hat{y}_1^l)~\ldots ~(\hat{s}_{m-1}^l,\hat{y}_{m-1}^l)~(\hat{s}_{m-1}^r,\hat{y}_{m-1}^r)$, where $\hat{y}_{j}^l$ and $\hat{y}_{j}^r$ are the values that the quantized function assumes in the extreme points of the segment $S_j$ and, again, $(\hat{s}_{j}^r,\hat{y}_{j}^r)=(\hat{s}_{j+1}^l,\hat{y}_{j+1}^l)$.
With these assumptions, in each segment $j$, we have $m_j=\frac{\hat{y}_j^r-\hat{y}_j^l}{\hat{s}_j^r-\hat{s}_j^l}$ and $q_j=\hat{y}_j^l$.  
	\subsubsection{Polynomial approximation}\mbox{} \\*
		In order to improve the accuracy of the approximation within each interval and consequently reduce the number of segments required to obtain a given precision, we can use a polynomial approximation.
Without imposing any continuity constraint across the intervals, and given the polynomial degree $d$, the coefficients of the polynomial that minimizes $\max_{\hat{x}\in S_j}\{|\hat{f}(\hat{x})-\sum_{i=0}^d a_{j,i}(\hat{x}-\hat{s}^l_j)^i|\}$ in a generic interval $j$ can be obtained by searching the coefficients of the regression line that better approximates the set of points $\left\{\left(1,~ \hat{x},~ \hat{x}^2,~ \ldots,~ \hat{x}^d, \hat{f}(\hat{x})\right)\right\}_{\hat{x}\in S_j}$ in a space of $d+2$ dimensions.
If the approximation error given by the polynomial is not lower than the threshold, the interval is split again and a new polynomial is searched in the new partition.

\subsubsection{Continuous polynomial approximation}\mbox{} \\*
If we require that the approximation is continuous on the border of different segments, we can impose that the values assumed by the polynomial on the extreme points of the segments are equal to those assumed by the to-be-approximated function.  The polynomial of degree $d$ approximating $\hat{f}()$ in a section $S_j$, can then be obtained  by the polynomial of degree $d-1$ approximating the function.

Starting from a linear approximation in the interval obtained as described in \sect{sec:cla}, a quadratic approximation function can be obtained as $\widetilde{f}^2_j(\hat{x})=b_{j,0}+b_{j,1}(\hat{x}-\hat{s}^l_j)+b_{j,2}(\hat{x}-\hat{s}^l_j)(\hat{x}-\hat{s}^r_j)$, where $b_{j,0}=q_j$ and $b_{j,1}=m_j$. 
The approximation error inside the interval depends on the value of $b_{j,2}$ and can be expressed as 
\begin{eqnarray}
&&\epsilon_2(\hat{x},b_{j,2})=\nonumber\\
&=&\hat{f}(\hat{x})-b_{j,0}-b_{j,1}(\hat{x}-\hat{s}^l_j)-b_{j,2}(\hat{x}-\hat{s}^l_j)(\hat{x}-\hat{s}^r_j)=\nonumber\\
&=&\epsilon_1(\hat{x})-b_{j,2}(\hat{x}-\hat{s}^l_j)(\hat{x}-\hat{s}^r_j),
\end{eqnarray} 
where  $\epsilon_1(\hat{x})$ is the approximation error introduced by the linear approximation.
To obtain $b_{j,2}$, the mean square error $\frac{1}{|S_j|}\sum_{\hat{x}\in S_j}\epsilon_2^2(\hat{x})$ is minimized by imposing that its derivative with respect to $b_{j,2}$ is equal to 0. Doing so, we obtain 
\begin{equation}
b_{j,2}=\frac{\sum_{\hat{x}\in S_j}\epsilon_1(\hat{x})(\hat{x}-\hat{s}^l_j)(\hat{x}-\hat{s}^r_j)}{\sum_{\hat{x}\in S_j}(\hat{x}-\hat{s}^l_j)^2(\hat{x}-\hat{s}^r_j)^2}.
\end{equation}
Being interested to obtain an approximation of the form $\sum_{i=0}^d a_{j,i}(\hat{x}-\hat{s}^l_j)^i$, we find the following coefficients: $a_{j,0}=b_{j,0}$, $a_{j,1}=b_{j,1}-a_{j,2}(\hat{s}^r_j-\hat{s}^l_j)$ and $a_{j,2}=b_{j,2}$.

Given the approximation of degree 2, we can obtain the approximation of degree 3 similarly, as $\widetilde{f}^3_j(\hat{x})=\widetilde{f}^2_j(\hat{x})+b_{j,3}(\hat{x}-\hat{s}^l_j)(\hat{x}-\hat{s}^r_j)(\hat{x}-c_3)$, where $c_3$ is a point of the interval 
(possibly one of the extreme points), whose selection can be carried out by means of numerical analysis \cite{hamming2012numerical}.
By iterating the above operation, we can obtain an approximation of degree $d$, that can be written in the form $\sum_{i=0}^d a_{j,i}(\hat{x}-\hat{s}^l_j)^i$. 

\subsubsection{Other solutions}\mbox{} \\*
		Other techniques can be used to approximate $\hat{f}()$ within the segments $S_j$. For instance, by using a spline interpolation of degree $d$ we would ensure the continuity of the piecewise approximation and its first $d-1$ derivatives \cite{wahba1990spline}, however the spline approximation does not depend on the values that the function assumes in the non-extreme points of the segment. This raises some problems with the overall interpolation procedure. Suppose, for instance that at a certain point, with the domain split into $N$ segments, the maximum approximation error exceeds the desired maxim value. With a spline approximation is not easy to decide which segment should be further split, since even by splitting the segments where the error exceeds the threshold, it is possible that the new spline approximation exceeds the maximum error in some of the segments that have not been split. 

Alternatively, we could use a Taylor approximation centered in the middle of each interval. Unluckily, this solution provides an excellent approximation close to the center of the intervals, but deteriorates rapidly towards the extreme points. To solve the problem, polynomials of large degree should be used, otherwise we risk to split the domain into too many small intervals, making the solution inefficient.

As an additional possibility, we mention the usage of a neural networks, whose implementation in the encrypted domain has been proposed in  \cite{barni2011privacyECG}.
In fact the multi-layer perceptron (MLP) is a universal function approximator, as proven by the universal approximation theorem \cite{csaji2001approximation}. However, the proof is not constructive regarding the number of neurons required or the settings of the weights. Moreover each neuron involves several products, which can not be implemented efficiently in the encrypted domain, and an activation function, whose best secure implementation so far is based on a linear piecewise approximation.

 \section{Parameter representation}\label{sec:error}
	
In this section we evaluate the impact that the number of bits used to represent the parameters of the approximating function has on the approximation accuracy.

As we said, SMPC works with integer numbers, however the coefficients of the polynomials derived in the previous section are real numbers and need to be approximated with integer numbers. A possibility would be to simply approximate them by using the formula $\sum_{i=0}^d\lfloor a_{j,i}\rceil (\hat{x}-\hat{s}_j^l)^i$. Such a choice, however, may result in an exceedingly large approximation error.
To alleviate this problem, we quantize the approximation coefficients by multiplying them by a factor $k$ and dividing the final approximation result by the same value.
We underline that by imposing that $k$ is a power of 2, we can implement the division very easily by discarding the $\ell_k$ least significant bits of the result.
We also argue that a different number of bits (a different precision) is needed to represent the coefficients of different orders in the polynomial (intuitively more bits will be needed for higher orders). To allow for such a differentiation we introduce different multipliers for different coefficients, let us denote them by $k_{i}=2^{\ell_{k,i}}\leq k$, where $\ell_{k,i}$ is the number of bits used to represent the fractional part of $i$-th order coefficients, obviously such parameters have to be scaled during the computation so that all the parameters are amplified by the same factor $k$.

At this point we need to understand how many bits are needed to represent the coefficients. 
Since the coefficients can  be negative, we need one bit for the sign.
The number of bits used to represent the magnitude can change with the coefficient degree, but its must be the same for each interval.
The magnitude of the parameters of degree $i$ depends first of all on the biggest value assumed by $a_{j,i}$ for all the sections $S_j$, identified as $\ell_{u,i}=\lceil\log_2(\max_j\{a_{j,i}\})\rceil$. In such a way a different bitsize is used for coefficients corresponding to different degrees in the polynomial.
Then the magnitude depends on the quantization factor $k_i$. This is equal to understand how many bits of the fractional part of the parameters $a_{j,i}$ are represented for each degree $i$. Considering that the largest quantizer for the parameters is $k_d=k$, we can rewrite the approximating function by using the notation introduced so far as\footnote{The formula refers to the $j$-th segment for simplicity.}
\begin{equation}
\tilde{f}(\hat{x})=\left\lfloor \frac{\sum_{i=0}^d2^{\ell_k-\ell_{k,i}}\lfloor k_i a_{j,i}\rceil (\hat{x}-\hat{s}_j^l)^i}{k} \right\rfloor, \label{eq:quantizedbis}
\end{equation}
allowing us to evaluate products involving values represented with smaller bit-lengths and then multiplying the results by the factor $k/k_i=2^{\ell_k-\ell_{k,i}}$ by simply concatenating a proper number of zeroes. 
While $k_i a_{j,i}$ can be represented as an integer number by rounding it in the plain domain, the division in (\ref{eq:quantizedbis}) is computed discarding the $\log_2 k=\ell_k$ less significant bits of the sum result, allowing only to truncate the value.

To determine $\ell_{k,i}$, we compute the difference between the quantized function $\hat{f}(\hat{x})$ and the approximating function $\tilde{f}(\hat{x})$. Considering that the analysis is the same for each segment $j$, in the following we omit such an index for simplicity. The approximation error can be written as:
\begin{align}
&\hat{f}(\hat{x})-\tilde{f}(\hat{x})=\nonumber\\
&=\hat{f}(\hat{x})-\left\lfloor \frac{\sum_{i=0}^d2^{\ell_k-\ell_{k,i}}\lfloor k_i a_{i}\rceil (\hat{x}-\hat{s}^l)^i}{k} \right\rfloor=\nonumber\\
&=\hat{f}(\hat{x})-\left( \frac{\sum_{i=0}^d2^{\ell_k-\ell_{k,i}} (k_i a_{i}+\epsilon_i) (\hat{x}-\hat{s}^l)^i}{k} +\epsilon_t\right)=\nonumber\\
&=\hat{f}(\hat{x})-\sum_{i=0}^d a_i(\hat{x}-\hat{s}^l)^i - \left(\sum_{i=0}^d \frac{\epsilon_i(\hat{x}-\hat{s}^l)^i}{k_i}+\epsilon_t\right),
\end{align}

where $0\leq\epsilon_t<1$ is the truncation error, while $|\epsilon_i|<1/2$ is the error introduced to round the $i$-th coefficient. 
$\hat{f}(\hat{x})-\sum_{i=0}^d a_i(\hat{x}-\hat{s}^l)^i$ is the approximation error, while $\sum_{i=0}^d \frac{\epsilon_i(\hat{x}-\hat{s}^l)^i}{k_i}+\epsilon_t$ is an additional representation error that we want to keep as small as possible. Since $\epsilon_t<1$ (equivalent to the codomain quantization error), we impose that $\sum_{i=0}^d \frac{\epsilon_i(\hat{x}-\hat{s}^l)^i}{k_i}$ is also lower than one, so that the representation error is lower than twice the codomain quantization step. 
Recalling that each interval contains a number of points that is a power of 2, the difference between the input and the left extreme of the segment the input belongs to is $\hat{x}-\hat{s}^l<2^{\ell_v}\triangleq\max_j\{\hat{s}_j^r-\hat{s}_j^l\}$. Considering that $|\epsilon_i|<1/2$, we obtain 
\begin{equation}
\left|\sum_{i=0}^d \frac{\epsilon_i(\hat{x}-\hat{s}^l)^i}{k_i}\right|<\sum_{i=0}^d \frac{2^{i\ell_v}}{2k_i}.
\end{equation}
Allowing an error lower than $1/(d+1)$ in each term of the sum ensures that the total error is lower than 1, yielding $k_i>(d+1)2^{i\ell_v-1}$.

The bitsize of the fractional part of the $i$-th coefficient then is $\ell_{k,i}=i\ell_v+\lceil\log_2 (d+1)\rceil-1$ and the bitsize of the amplification factor $k$ can be obtained by letting $k=k_d$.

The total number of bits needed to represent all the parameters of the approximation ($\lfloor k_ia_{j,i}\rceil ~ \forall i$ and $\hat{s}^l_j$) then is $\ell_p=\ell_x+\sum_{i=0}^d (\ell_{u,i}+\ell_{k,i}+1)$.

 \section{STPC implementations}\label{sec:protocols}
   As already outlined in Section \ref{sec:tools}, the main tools for STPC are HE and GC. Moreover, it is also possible to develop hybrid protocols by composing several subprotocols, each one implemented by relying on the most suitable approach.

In the following, we present two solutions to implement the approximation algorithm described in the previous sections in a STPC setting. The two solutions rely, respectively, on a full-GC implementation and a new hybrid solution. We excluded a priori the development of a protocol entirely based on HE since the protocols introduced in Section \ref{sec:approximation} need the bit decomposition of the input, for which an efficient HE implementation does not exist. 

We first describe the two protocols and then we analyze their complexity in \sect{sec:comparing} to evaluate which of the two is more efficient for different setups.
We assume that the input $\hat{x}$ is available to Alice in form of garbled secrets, while
the full-GC and hybrid protocols represent the output value $\hat{y}$ as garbled secrets and as a cyphertext respectively. If a different representation is needed for further computation, the conversion algorithm presented in \cite{kolesnikovdust} must be used. The above setting mimics a case in which the function evaluation protocol is embedded inside an outer protocol. In addition of being used for further computation, the output can also be disclosed to Alice as a final result, while $\hat{x}$ cannot be an input from Alice or Bob, otherwise, being the function known, they could directly input $f(\hat{x})$ to the subsequent computation.

\subsection{Full-GC solution}\label{sec:GCimplementation}
Given an approximating function having the form defined in the previous section, its evaluation in correspondence of an input $\hat{x}$ can be implemented in three steps: i) identification of the segment $\hat{x}$ belongs to; ii) retrieval of the parameters determining the approximating function in the identified segment;  iii) use of the retrieved parameters to compute $\hat{f}(\hat{x})$. In this section, we present a protocol entirely based on GC to implement the above steps.

\subsubsection{Interval detection}  \label{sec:GCprimo} 
\begin{figure}[!hbt]
\centering
\ifdraft
\includegraphics[width=.5\columnwidth]{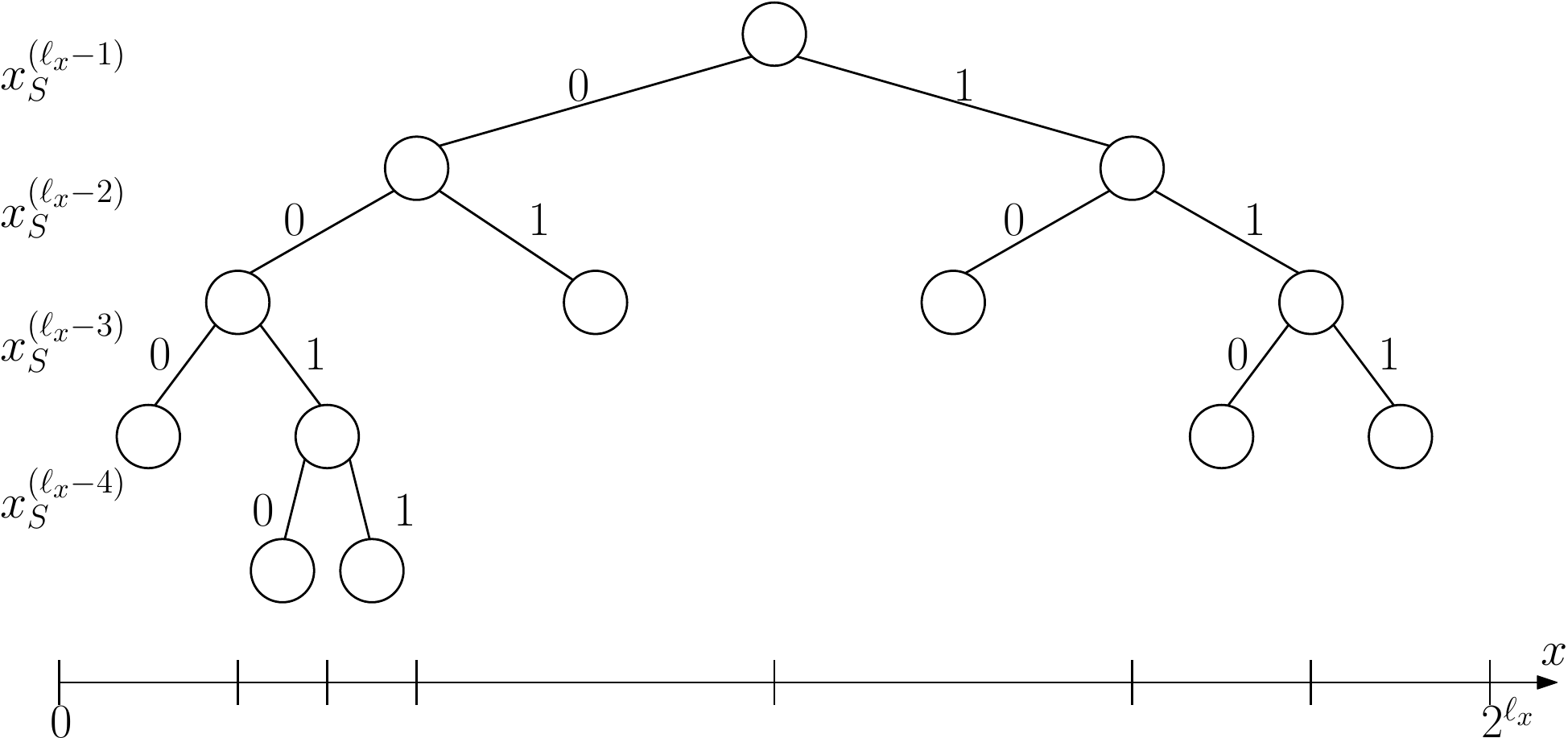}
\else
\includegraphics[width=\columnwidth]{bisection}
\fi
\caption{Example of binary tree associated to a domain partition.}
\label{fig:selectiontree}
\end{figure}

The correct subset is identified through a classification tree.
Thanks to the use of a bisection algorithm we can associate a binary tree to the partition, where we can reach the leaf associated to the segment the input belongs to by traversing the tree from the root node and by choosing the left or right child according to the most significant bit of the input, and repeating the operation for each node according to the next bits, until a leaf is reached, as shown in \fig{fig:selectiontree}.

The idea behind the GC protocol is to build a binary circuit that, given the input, returns as many output bits as the number of leaves in the tree and where only the bit corresponding to the leaf associated to the correct segment is equal to 1.

Let us assume that the binary tree has $N$ leaves (and hence $N-1$ nodes).
An input $\hat{x}$ belongs to the $j$-th subset $S_j$ (associated to the $j$-th leaf of the tree) if $p$ is the depth of the leaf in the tree and the $p$ most significant bits of $\hat{x}$ and the left extreme $s^l_j$ are equal. These $p$ bits are associated to the direct path  from the root to the leaf associated to segment $S_j$, i.e. the sequence of bits encountered on the branches when traveling from the root to the leaf (\fig{fig:selectiontree}). 

The above condition can be verified through a binary circuit computing $\wedge_{i=1}^p \overline{\hat{x}_{\langle\ell_x-i\rangle}\xor {s}^l_{j\langle\ell_x-i\rangle}}=1$, where $\wedge$ indicates the AND operator, $\xor$ the XOR operator, ~$\bar{}$~ the negation and $\hat{x}_{\langle i\rangle}$ identifies the $i$-th bit of $\hat{x}$ ($\hat{x}_{\langle 0\rangle}$ is the least significant bit). 

Being the path from the root to each leaf known, $\overline{\hat{x}_{\langle\ell_x-i\rangle}\xor {s}^l_{j\langle\ell_x-i\rangle}}$ can be simply evaluated as $\overline{\hat{x}_{\langle\ell_x-i\rangle}}$ if ${s}^l_{j\langle\ell_x-i\rangle}=0$ and as $\hat{x}_{\langle\ell_x-i\rangle}$ otherwise, for each $i$.
The AND gates common to different paths can be evaluated only once. During the tree design, each time a node is added, except for the root node, two AND gates are added: one having as inputs the upper path and the negation of the actual bit of $\hat{x}$, the other with the upper path and the actual bit of $\hat{x}$. To decrease the total number of gates the first AND gate and the NOT gate between the input bit and the AND can be merged together and replaced by a gate with the following truth table 
\begin{center}
{\small
\begin{tabular}{cc|cl}
1st input & 2nd input & output&\\
\hline
0&0&0&\\
0&1&0&\\
1&0&1&\\
1&1&0&.\\
\end{tabular}}
\end{center}

\begin{figure}[!hbt ]
\centering
\ifdraft
\includegraphics[width=.5\columnwidth]{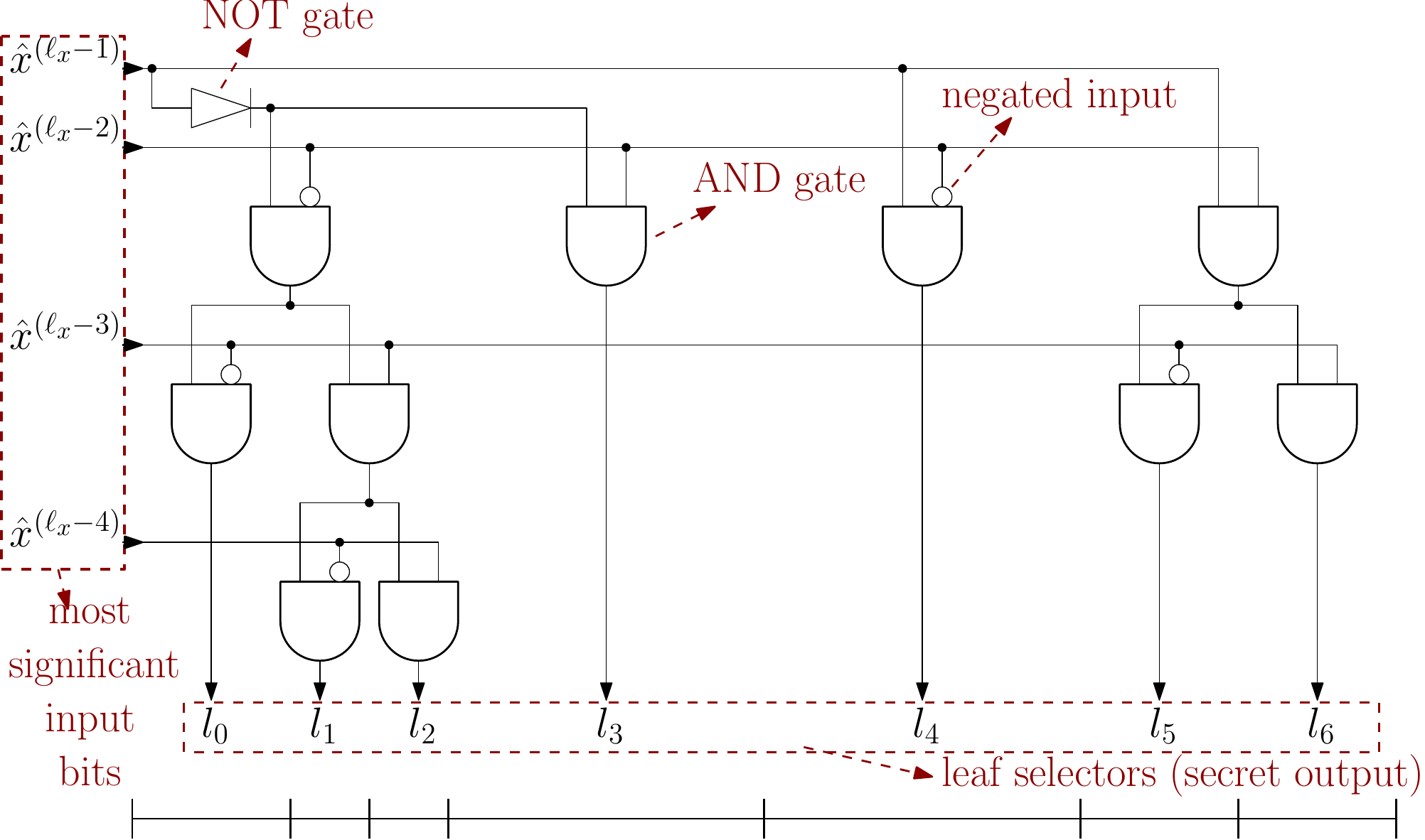}
\else
\includegraphics[width=\columnwidth]{circuit1}
\fi
\caption{Circuit implementing the subset identification associated to the binary tree of \fig{fig:selectiontree}.}
\label{fig:tree_circuit}
\end{figure}

The final circuit implementing the bisection algorithm is composed by only $2(N-2)$ non-XOR gates (an example is shown in \fig{fig:tree_circuit}, where the circuit implementing the bisection algorithm depicted in figure \ref{fig:selectiontree} is shown),  significantly improving the circuit used in the protocol described in \cite{pignata2012general}, where segment selection was achieved through $N-1$ comparison circuits, and required $(N-1)\ell_x$ non-XOR gates. We point out that such an improvement was possible due to the use of a bisection algorithm during the construction of the approximating function.

\subsubsection{Parameters selection} \label{sec:GCsecondo}
\begin{figure}[hbt!]
\centering
\ifdraft
\includegraphics[width=.45\columnwidth]{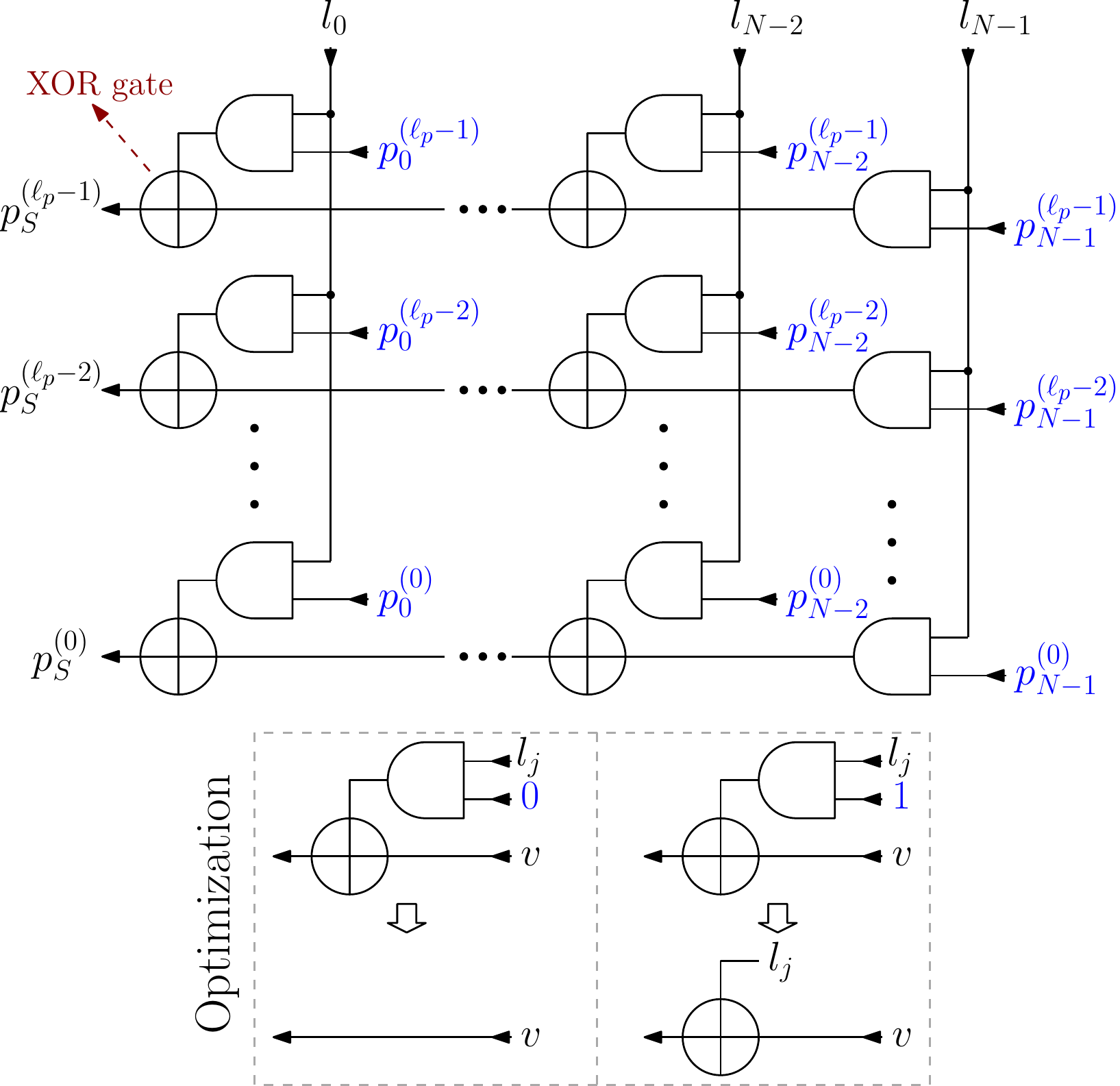}
\else
\includegraphics[width=.9\columnwidth]{circuit2}
\fi
\caption{Circuit for parameters selection and its optimizations (blue inputs are public).}
\label{fig:parameter_selection}
\end{figure}
The circuit for parameters selection is entirely composed by XOR gates.

Given the concatenation $p_i$ of the approximation parameters associated to each segment $i$, and the output of the associated leaf $l_i$, and by remembering that if $\hat{x}$ belongs to the segment $S_j$ only $l_{j}=1$,
while the outputs associated to the other leaves are null, the parameters $p_j$ of the segment $S_j$ can be obtained as $p_j=\sum_{i=1}^N l_i\, p_i$. Considering that only $l_j\, p_j\neq 0$, the same result can be obtained by evaluating $\xor_{i=1}^N l_i\, p_i$. Finally we recall that the to-be-approximated function is public, then the domain partition and the parameters of the various segments are not secret, hence the AND between the output of any leaf $i$ 
and a generic bit $b$ of the associated parameters $p_i$ can be expressed as 
\begin{equation}
l_i\, p_{i\langle b\rangle}=\left\{\begin{array}{lcl}
0 && \mbox{if }p_{i\langle b\rangle}=0\\
l_i && \mbox{otherwise.}
\end{array}\right.
\end{equation}
When $l_i\, p_{i\langle b\rangle}=0$, the XOR operation is implemented by simply propagating the other input (even if XOR gates have negligible complexity, in this way we further reduce the circuit complexity). An overall sketch of the circuit for parameters selection is given in figure \fig{fig:parameter_selection}.

We conclude by stressing out again that non-XOR gates are not used in the circuit and that less than $(N-1)\ell_p$ XOR gates compose the circuit. This marks a significant improvement with respect to the solution proposed in \cite{pignata2012general} where $N-1$ multiplexers are used, for a total of $(N-1)\ell_p$ non-XOR gates.

\subsubsection{Approximation}\label{sec:GCterzo}

During the last step, the approximation coefficients obtained in the previous phase are used to compute the approximated value. In the case of constant approximation no further operation is needed, since the approximation coincides with the parameters obtained in the second step.
This is not the case for linear and polynomial approximations. Let us assume, then, that a polynomial of degree $d$ is used for the approximation, and let us indicate the parameters determining the exact form of the approximating polynomial with $p_j$, while the secrets relative to $\hat{x}$ are available since the very beginning of the protocol.

First of all the difference $\delta$ between the input $\hat{x}$ and the left extreme $\hat{s}^l_j$ of the segment $S_j$ containing $\hat{x}$ is computed by a subtraction circuit composed by $\ell_x$ non-XOR gates \cite{lazzeretti2012thesis}. The sign of the output is discarded, since we know that the difference is always positive.
The direct evaluation of the polynomial $2^{\ell_k-\ell_{k,0}}\lfloor k_0a_{j,0}\rceil+\sum_{i=1}^d 2^{\ell_k-\ell_{k,i}}\lfloor k_ia_{j,i}\rceil\delta^i$ (extension of the solution proposed in \cite{pignata2012general} to polynomial of degree $d$) would require $d-1$ products with increasing complexity to compute all the powers of $\delta$ and then other $d$ products to multiply them with the corresponding coefficients. This would require a circuit composed by $\mathcal{O}(d^3\ell_v^2)$ non-XOR gates. In fact $\delta^i$ is represented by $i\ell_v$ bits, while the corresponding coefficient by $\ell_y+i\ell_v$ bits, hence their product needs $\mathcal{O}(i^2\ell_v^2)$ non-XOR gates. Summing the non-XOR gates of the $d$ products, we obtain a circuit composed by $\mathcal{O}(d^3\ell_v^2)$ non-XOR gates.

\begin{figure}[!hbt]
\centering
\ifdraft
\includegraphics[width=.35\columnwidth]{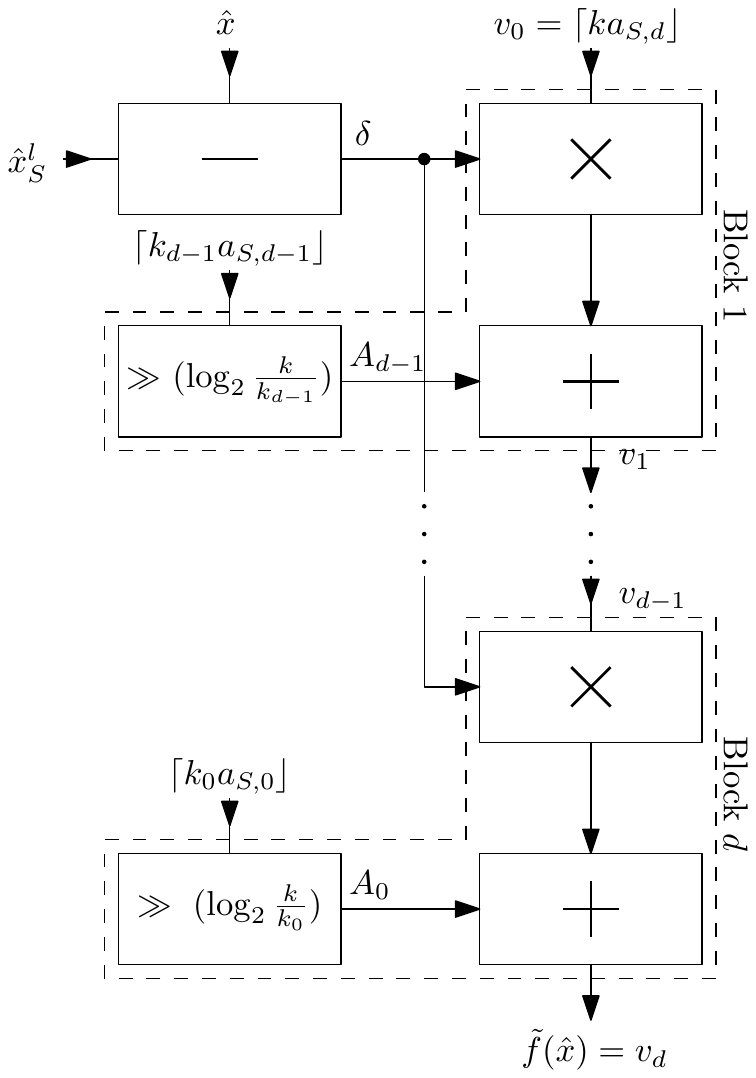}
\else
\includegraphics[width=.63\columnwidth]{circuit3}
\fi
\caption{Full-GC implementation of the approximation part.}
\label{fig:GCapprox}
\end{figure}

A simpler solution can be obtained by evaluating the approximating polynomial through a sequence of $d$ blocks, as shown in \fig{fig:GCapprox}. In practice the polynomial is evaluated as $A_0+x(A_1+x(A_2+x(\ldots +x(A_{d-1}+xA_d)\ldots)))$, where $A_i=2^{\ell_k-\ell_{k,i}}\lfloor k_ia_{S,i}\rceil$. By setting
$v_0=\lfloor ka_{S,d}\rceil$ and evaluating $d$ linear expressions
$$v_i=A_{d-i} + \delta v_{i-1},$$ (one for each block) with $i$ ranging from 1 to $d$, where $v_i$ is the output of the $i$-th block used as input to the next one. 

%
%
%

In the first product of the series, we compute the product between $v_0$, represented with $\ell_v$ bits, and the corresponding coefficient, represented with $d\ell_v+\ell_y$ bits, hence the product needs $2(d\ell_v+\ell_y)\ell_v-(d\ell_v+\ell_y)$ non-XOR gates and returns a value that is represented with $(d+1)\ell_v+\ell_y$ bits. The result is used in an ADDER together with the scaled coefficient, whose $\ell_v$ least significant bits are 0, hence an ADDER having $(d+1)\ell_v+\ell_y-\ell_v=d\ell_v+\ell_y$ non-XOR gates is used and its output is represented with $(d+1)\ell_v+\ell_y$ bits.
In a generic step $i$, the product is composed by $2((d+i-1)\ell_v+\ell_y)\ell_v-((d+i-i)\ell_v+\ell_y)$ non-XOR gates, its output is represented by $(d+i)\ell_v+\ell_y$ bits and hence the ADDER should be composed by $(d+i)\ell_v+\ell_y$ non-XOR gates, but considering that the $i\ell_v$ least significant bits of the parameter $a_i$ are null, only $d\ell_v+\ell_y$ non-XOR gates are needed.
In total the number of non-XOR gates composing the circuit computing the approximation is 
\begin{equation} 
3d^2\ell_v^2-d\ell_v^2+2d\ell_v\ell_y-1/2d^2\ell_v+1/2d\ell_v,\label{eq:gates3}
\end{equation} 
for a complexity that is reduced to $\mathcal{O}(d^2\ell_v^2)$. 


\subsection{Hybrid solution}\label{sec:Hybridimplementation}
The second protocol that we developed relies on a hybrid combination of GC and additively homomorphic encryption. The rational behind it is that while the first two steps of the protocol are more easily carried out by resorting to GC, for the last one the use of HE could be advantageous. In fact, an HE implementation of the interval detection requires the encryption of single bits, resulting in a high expansion factor, and the implementation of boolean gates through more expensive products and the number of exponentiations between cyphertexts, while parameter section is evaluated essentially \emph{for free} by using GC and hence any HE implementation would have a higher complexity.

A hybrid approach had already been used in \cite{pignata2012general}, where the input $\hat{x}$ is used to compute the linear approximation within all the segments through HE and then the bits obtained by comparing the input with the left extremes of the intervals are obfuscated and used to select the correct approximation, by using an interactive HE protocol. Such a  protocol can be easily extended to a generic polynomial approximation.

Here we propose to use an approach similar to the one presented in \sect{sec:GCsecondo}: we first select the parameters with negligible complexity and then pass them to the multipliers rather then computing all the multiplications and then select the correct one, avoiding $N-1$ multiplexers and the transmission of $\mathcal{O}(N)$ cyphertexts. 
Hence, in contrast to \cite{pignata2012general}, we introduce a new hybrid protocol for which  the complexity of the homomorphic part does not depend on the number of intervals the domain is partitioned into, but only on the degree of the polynomial.

To describe the protocol, we assume that for the GC part Bob acts as the garbler and Alice as the evaluator, while in the HE part, Alice owns the private key while the public key is obviously available to Bob as well. We also assume that $\hat{x}$ is available to Alice through a previous computation in the form of garbled secrets. As we said, the segment $S_j$ with $\hat{x}$ and the corresponding approximation parameters are obtained by using GC as detailed in \sect{sec:GCprimo} and \sect{sec:GCsecondo}.
At this point, the parameters $\lfloor k_ia_{j,i}\rceil$ (in the next we avoid the round operator) are obfuscated by adding to them a value $r_{i}^{(a)}$ randomly generated by Bob.

Similarly to \sect{sec:GCterzo}, we assume that each parameter $k_ia_{S,i}$ is represented with $1+\ell_{u,i}+i\ell_v+\lfloor\log_2 (d+1)\rfloor$ bits, hence
the relative obfuscation value is  $1+\ell_{u,i}+i\ell_v+\tau$ bits long.
The difference $\delta$ between $\hat{x}$ and $\hat{s}^l_S$ is computed by using a subtraction circuit and the result is also obfuscated by adding a random number $r$ provided by Bob as an additional input of the GC and represented with $\ell_v+\tau$ bits.
The difference and the obfuscated parameters are sent to Alice.

Alice computes all the powers of $\delta+r$ from 1 to $d$, encrypts them and sends the cyphertexts to Bob.
Knowing $r$, Bob can remove the obfuscation by computing
\begin{eqnarray}
\enc{\delta}&=&\enc{\delta+r}\enc{-r}\nonumber\\
\enc{\delta^2}&=&\enc{(\delta+r)^2}\enc{\delta}^{-2r}\enc{-r^2}\nonumber\\
\enc{\delta^3}&=&\enc{(\delta+r)^3}\enc{\delta^2}^{-3r}\enc{\delta}^{-3r^2}\enc{-r^3}\nonumber\\
&&\dots\nonumber
\end{eqnarray}

To evaluate the polynomial, Bob needs again to interact with Alice, to compute the product between the powers of $\delta$ and the corresponding coefficients. Hence Bob introduces a new obfuscation $r^{(\delta)}_i$ in each power $\delta^i,~i=2,\ldots,d$ and sends the value to Alice. In this way Alice, after decryption, obtains all the powers of $\delta$ obfuscated by a value that Bob knows exactly (Alice already has $\delta$ obfuscated by $r^{(\delta)}_1=r$).

Alice computes the polynomial by using the obfuscated powers and parameters obtaining:
\begin{equation}
y_{ob}=
\frac{k}{k_0}(k_0a_{S,0}+r_0^{(a)})+ \sum_{i=0}^d\frac{k}{k_i}(k_ia_{S,i}+r_i^{(a)})(\delta^i+r^{(\delta)}_i)
\end{equation}
that is equal to the polynomial evaluated in $\hat{x}$ obfuscated by the value
\begin{equation}
\frac{k}{k_0}r_0^{(a)}+\sum_{i=1}^d \frac{k}{k_0}r_i^{(a)}\delta^i+\sum_{i=1}^dr^{(\delta)}_i\frac{k}{k_i}(k_ia_{S,i}+r_i^{(a)}),
\end{equation}
where we recall that $\frac{k}{k_i}=2^{\ell_k-\ell_{k,i}}$.

Alice encrypts $y_{ob}$ and all the quantities $-\frac{k}{k_i}(k_ia_{S,i}+r_i^{(a)})$ and sends them to Bob that finally computes  under encryption
{\small
\begin{equation}
\enc{\hat{y}}=\enc{y_{ob}}\enc{-\frac{k}{k_0}r_0^{(a)}} \prod_{i=1}^d \enc{\delta^i}^{-\frac{k}{k_0}r_i^{(a)}} \prod_{i=1}^d \enc{-\frac{k}{k_i}(k_ia_{S,i}+r_i^{(a)})}^{r^{(\delta)}_i}.
\end{equation}}

It is important to underline that when $d=1$, the first part of the protocol is not needed, because Bob already knows the obfuscation affecting the difference, hence the protocol starts directly with the computation of $y_\mathit{ob}$.

In contrast to the full-GC solution, the hybrid protocol outputs the value amplified by a factor $k$ and enriched with many fractional bits. If it is necessary to represent the final value with only $\ell_y$ bits for further computation, we can resort to one of the interactive protocols proposed in \cite{veugen2010encrypted,lazzeretti2011division}.

 \section{Protocols comparison}\label{sec:comparing}
    In this section we evaluate the efficiency of the two protocols described so far, by considering both communication and computational complexity. We also provide the runtime of a Java implementation of the protocols.

As shown in \sect{sec:error}, the parameters associated to higher degree approximations require more bits for their representation. Hence the use of polynomials with a large degree is convenient only if it permits to significantly decrease the number of segments the domain is partitioned into. For this reason the results strongly depend on the function to be approximated. Generally speaking, the most efficient solution must be decided by considering the specific applications and the available hardware/software. Here, we exemplify the kind of analysis that the designer should carry out, by considering the approximation of the function $\mathit{sinc}(x)=\frac{\sin(\pi x)}{\pi x}$ in the interval $[0,10)$.

For sake of simplicity,  we assume that $\ell_x=\ell_y=\ell$, hence the parameters influencing the complexity are the bitlength $\ell$, the polynomial degree $d$ and the number of segments $N$ used to partition the function domain. In addition, we must consider the security parameters involved in the protocols and summarized in  \tab{tab:parameters}.
\begin{table}[!t]
\begin{center}
\caption{Short term security parameters \cite{girycryptographic}.}
\label{tab:parameters}
\footnotesize
\begin{tabular}{lcc}
\hline
Content  & Name & Bitsize \\
\hline\hline
Homomorphic security parameter& $T$ & $1024$\\
Garbled circuit security parameter& $t$ & $80$\\
Obfuscation parameter for HE to GC conversion & $\tau$ & $80$\\
\hline
\end{tabular}
\end{center}
\end{table}

First of all we investigate the effect that the polynomial degree has on the number of segments of the partition. 
Given the bitlength $\ell$, the function is first scaled and translated to fit the domain $[0,2^\ell)$ and the codomain $[0,2^\ell)$. We report in \tab{tab:sections} the number of segments obtained by using constant, linear, quadratic and cubic approximation as a function of the bitlength $\ell$ and the target approximation error $\epsilon$ (computed as $\mathit{error}/\max{(y)}$). We can easily observe that by using a cubic approximation (or even higher degrees) the number of segments continues to decrease, but no significant improvements are obtained, hence the cubic solution will not be considered any further.
\begin{table}[t]
\begin{center}
\caption{Number of segments required to approximate $\mathit{sinc}(x)$ as a function of the bitlength $\ell$ and the approximation error $\epsilon$ (missing values indicate that the given error cannot be reached with the given bitlength). }
\label{tab:sections}
\begin{tabular}{ccrrrrrr}
\multicolumn{8}{c}{(a) Constant approximation}\\
\hline
$\ell \backslash\epsilon$&& 0.1 &0.05& 0.01& 0.005& 0.001& 0.0005\\
\hline\hline
         8 &            &         13 &         28 &         92 &        127 &           &          \\
        12 &            &         15 &         33 &        171 &        313 &       1158 &       1998 \\
        16 &            &         15 &         33 &        182 &        361 &       1724 &       3408 \\
        20 &            &         15 &         35 &        184 &        365 &       1743 &       3482 \\
\hline
\multicolumn{8}{c}{}\\
\multicolumn{8}{c}{(b) Linear approximation}\\
\hline
$\ell \backslash\epsilon$&& 0.1 &0.05& 0.01& 0.005& 0.001& 0.0005\\
\hline\hline
         8 &            &          8 &         17 &         36 &         49 &           &           \\
        12 &            &          7 &         17 &         38 &         53 &        124 &        214 \\
        16 &            &          7 &         17 &         38 &         53 &        120 &        166 \\
        20 &            &          7 &         17 &         38 &         53 &        119 &        162 \\
\hline
\multicolumn{8}{c}{}\\
\multicolumn{8}{c}{(c) Quadratic approximation}\\
\hline
$\ell \backslash\epsilon$&& 0.1 &0.05& 0.01& 0.005& 0.001& 0.0005\\
\hline\hline
         8 &            &          5 &          9 &         16 &         22 &           &           \\
        12 &            &          5 &         10 &         17 &         21 &         35 &         66 \\
        16 &            &          6 &         10 &         17 &         21 &         35 &         43 \\
        20 &            &          9 &         10 &         17 &         21 &         35 &         42 \\
\hline
\multicolumn{8}{c}{}\\
\multicolumn{8}{c}{(c) Cubic approximation}\\
\hline
$\ell \backslash\epsilon$&& 0.1 &0.05& 0.01& 0.005& 0.001& 0.0005\\
\hline\hline
         8 &            &          5 &          7 &         16 &         23 &           &           \\
        12 &            &          5 &         7 &         16 &         21 &         34 &         52 \\
        16 &            &          5 &         7 &         16 &         21 &         35 &         37 \\
        20 &            &          5 &         10 &         16 &         21 &         35 &         37 \\
\hline
\end{tabular}
\end{center}
\end{table}
Given the number of segments, we can evaluate the communication complexity, the computational complexity and the runtime for the three kinds of approximations.
We remind that with regard to the piecewise constant approximation, the parameters provided by the first two steps of the protocol implemented entirely by mens of GC already represent the approximation we look for without the need of any further computation. For this reason, the use of a Hybrid protocol for the piecewise constant approximation is not necessary.

For sake of brevity continuous approximations are not considered, however we underline that their complexities are similar to those of the corresponding non continuous approximations.

\subsection{Communication complexity}

For the GC part we use precomputation only for the oblivious transfers, so that an $OT^1_2$ used to associate a $t$-bit secret to an input bit provided by Alice  requires the online transmission of $\sim 2t$ bits, while the transmission of the secrets associated to the input bits of Bob requires $t$ bits and for each non-XOR gate $3t$ bits are transferred, while no communication is needed for the XOR gates.
If circuit transmission is precomputed, as in the case of two parties knowing in advance that they have to evaluate together a given function, the communication complexity is reduced to only  the online transmission of the $\sim 2t$ bits relative to Alice's input.

\subsubsection{Full-GC solution} By assuming that the secrets associated to $\hat{x}$ are already available, the protocol requires only the transmission of the gates composing the circuit in one communication round.
As already shown in \sect{sec:GCimplementation}, $2(N-2)$ non-XOR gates are used in the first sub-circuit and 0 in the second sub-circuit, while the complexity of the third part depends on the polynomial degree, for which the  estimated number of gates is given by equation \eqref{eq:gates3}.
Increasing the polynomial degree decreases the number of segments, and hence the number of non-XOR gates in the first part of the circuit, especially when passing from constant to linear approximation with high $\ell$ and small $\epsilon$. On the other side, the number of non-XOR gates of the third part is $\mathcal{O}(d^2\ell^2)$ and hence its complexity increases.
\tab{tab:fullGCcommunication} shows the communication complexity of the full-GC protocol in bytes according to the results shown in \tab{tab:sections}, where the parameter bitlengths have been set according to the analysis carried out in \sect{sec:error}.
\begin{table}[t]
\begin{center}
\caption{Communication complexity of the online phase (in bytes) of the full-GC approximation protocol applied to $\mathit{sinc}(x)$ as a function of $\ell$ and $\epsilon$ (missing values mean that the given error cannot be reached with the given bitlength).}
\label{tab:fullGCcommunication}
\begin{tabular}{ccrrrrrr}
\multicolumn{8}{c}{(a) Constant approximation}\\
\hline
$\ell \backslash\epsilon$&& 0.1 &0.05& 0.01& 0.005& 0.001& 0.0005\\
\hline\hline
         8 &            &        660 &       1560 &       5400 &       7500 &    &     \\
        12 &            &        780 &       1860 &      10140 &      18660 &      69360 &     119760 \\
        16 &            &        780 &       1860 &      10800 &      21540 &     103320 &     204360 \\
        20 &            &        780 &       1980 &      10920 &      21780 &     104460 &     208800 \\
\hline
\multicolumn{8}{c}{}\\
\multicolumn{8}{c}{(b) Linear approximation}\\
\hline
$\ell \backslash\epsilon$&& 0.1 &0.05& 0.01& 0.005& 0.001& 0.0005\\
\hline\hline
         8 &            &       4320 &       3180 &       3870 &       4650 &            &            \\
        12 &            &       9180 &       7140 &       7710 &       8610 &      11790 &      17190 \\
        16 &            &      16020 &      13020 &      13350 &      14250 &      16710 &      19470 \\
        20 &            &      24780 &      20820 &      20910 &      21810 &      23730 &      26310 \\
\hline
\multicolumn{8}{c}{}\\
\multicolumn{8}{c}{(c) Quadratic approximation}\\
\hline
$\ell \backslash\epsilon$&& 0.1 &0.05& 0.01& 0.005& 0.001& 0.0005\\
\hline\hline
8 && 16620 & 22740 & 9000 & 9720 &  &  \\
12 && 39180 & 39420 & 22560 & 23160 & 19920 & 18120 \\
16 && 71400 & 71640 & 47520 & 48120 & 42240 & 43140 \\
20 && 91320 & 91500 & 82080 & 82680 & 74400 & 75240 \\
\hline
\end{tabular}
\end{center}
\end{table}

As it can be seen from the tables, constant approximation is preferable to linear approximation for small bitlengths with large representation error. Quadratic approximation has always a communication complexity larger than the linear approximation and the results worsen with higher polynomial degrees. It is important to underline that  the complexity depends more on the bitlength than on the precision. In fact, a smaller approximation error results in a smaller interval for each segment and hence the number of bits representing the parameters decreases, thus reducing the complexity of the third circuit as well.

\subsubsection{Hybrid solution}
The hybrid protocol requires 4 communication rounds. In the first one, the garbled circuit is transmitted, together with the secrets of the random values that are used to obfuscate the circuit outputs. As shown in \sect{sec:Hybridimplementation}, the circuit transmitted in the first round is composed by the $2(n-2)$ non-XOR gates composing the tree, the $\ell_v$ non-XOR gates of the subtraction circuit, $\ell_v+\tau$ bits for the adder used to obfuscate the difference  and $\ell_{u,i}+i\ell_v+\lfloor \log_2(d+1\rfloor)+\tau \simeq (i+1)\ell+\tau$ non-XOR gates for the obfuscation of each parameter $k_ia_i~\forall i=0\ldots d$.
In the second round $d$ cyphertexts, containing the powers of the obfuscated difference between the input value and the left extreme of the segment are transmitted,  while in the third round $d-1$ cyphertexts, containing the obfuscated powers, are sent back. Finally, in the fourth round, one cyphertext, the obfuscated approximation, and the $d$ cyphertexts encrypting the obfuscated parameters are sent to Bob.
When $d=1$, the second and third rounds are discarded and the cyphertext containing $\delta+r$ is transmitted during the last round.
The communication complexity of the linear and quadratic solutions are shown in \tab{tab:hyb1communication}.
\begin{table}[t]
\begin{center}
\caption{Communication complexity of the online phase (in bytes) of the hybrid approximation protocol applied to $\mathit{sinc}(x)$ as a function of $\ell$ and $\epsilon$ (missing values mean that the given error cannot be reached with the given bitlength). }
\label{tab:hyb1communication}
\begin{tabular}{ccrrrrrr}
\multicolumn{8}{c}{(a) Linear approximation [2 rounds]}\\
\hline
$\ell \backslash\epsilon$&& 0.1 &0.05& 0.01& 0.005& 0.001& 0.0005\\
\hline\hline
         8 &            &      11898 &      12218 &      13288 &      14068 &            &            \\
        12 &            &      12438 &      12818 &      14008 &      14908 &      19058 &      24458 \\
        16 &            &      13038 &      13418 &      14608 &      15508 &      19418 &      22178 \\
        20 &            &      13638 &      14018 &      15208 &      16108 &      19958 &      22538 \\
\hline
\multicolumn{8}{c}{}\\
\multicolumn{8}{c}{(b) Quadratic approximation [4 rounds]}\\
\hline
$\ell \backslash\epsilon$&& 0.1 &0.05& 0.01& 0.005& 0.001& 0.0005\\
\hline\hline
         8 &            &      16666 &      17346 &      17256 &      17976 &            &            \\
        12 &            &      17626 &      17986 &      18136 &      18736 &      19966 &      22036 \\
        16 &            &      18686 &      18966 &      19056 &      19656 &      20646 &      21546 \\
        20 &            &      19306 &      19526 &      19976 &      20576 &      21566 &      22406 \\
\hline
\end{tabular}
\end{center}
\end{table}
\tab{tab:choice} shows the best choice for each set of parameters, according to the communication complexity. In general, the full-GC solution is preferable, but for a large number of bits and high precision the hybrid protocol requires to transmit less data. It is interesting to observe that in a case ($\ell=20$ and $\epsilon=0.0005$) the hybrid quadratic approximation protocol provides slightly sbetter results than the others.
\begin{table}[!t]
\begin{center}
\caption{Best protocol for each configuration, according to communication complexity analysis (solution-degree). }
\label{tab:choice}
\begin{tabular}{ccrrrrrr}
\multicolumn{8}{c}{(a) Constant approximation}\\
\hline
$\ell \backslash\epsilon$&& 0.1 &0.05& 0.01& 0.005& 0.001& 0.0005\\
\hline\hline
 8 & &GC-0 &GC-0 &GC-1 &GC-1 &    &     \\
12 & &GC-0 &GC-0 &GC-1 &GC-1 &GC-1 &GC-1 \\
16 & &GC-0 &GC-0 &GC-0 &GC-1 &GC-1 &GC-1 \\
20 & &GC-0 &GC-0 &GC-0 &Hyb-1 &Hyb-1 &Hyb-2 \\
\hline
\end{tabular}
\end{center}
\end{table}

At least for the example discussed in this paper, if the circuit can be transmitted offline, hybrid solutions are in general not advantageous, and the choice among the different full-GC solutions depends only on computational complexity.

\subsection{Computational complexity}

As shown in \sect{sec:tools}, the computational complexity depends on the number of Hash  functions for the GC part of the protocols and the number of exponentiations for the HE part, while the XOR between secrets and products between cyphertexts have a negligible complexity.
We consider that garbling is performed online, otherwise the complexity is reduced only  to the online evaluation of the circuit (performed by Alice).

\subsubsection{Full-GC solution} As already said, the interval detection circuit consists of $2(N-2)$ non-XOR gates, while the number of non-XOR gates composing the interpolation circuit is given by equation \eqref{eq:gates3}.
\tab{tab:fullGCcomputation} shows the total computational complexity of the full-GC solution when constant, linear or quadratic approximation is used.
As for the communication complexity, the constant approximation is preferable to the linear approximation for small bit-lengths and large representation errors, while the quadratic approximation exhibits the worst performance. 

\begin{table}[t]
\begin{center}
\caption{Average computational complexity of the online phase expressed as number of Hash functions  (rounded to the closest integer) computed by Alice and Bob in the full-GC approximation protocol applied to $\mathit{sinc}(x)$. The complexity is expressed as a function of $\ell$ and $\epsilon$ (missing values mean that the given error cannot be reached with the given bitlength).}
\label{tab:fullGCcomputation}
\begin{tabular}{ccrrrrrr}

\multicolumn{8}{c}{(a) Constant approximation}\\
\hline
$\ell \backslash\epsilon$&& 0.1 &0.05& 0.01& 0.005& 0.001& 0.0005\\
\hline\hline
8 && 83 & 195 & 675 & 938 &  &  \\
12 && 98 & 233 & 1268 & 2333 & 8670 & 14970 \\
16 && 98 & 233 & 1350 & 2693 & 12915 & 25545 \\
20 && 98 & 248 & 1365 & 2723 & 13058 & 26100 \\
\hline
\multicolumn{8}{c}{}\\
\multicolumn{8}{c}{(b) Linear approximation}\\
\hline
$\ell \backslash\epsilon$&& 0.1 &0.05& 0.01& 0.005& 0.001& 0.0005\\
\hline\hline
8 && 540 & 398 & 484 & 581 &  &  \\
12 && 1148 & 893 & 964 & 1076 & 1474 & 2149 \\
16 && 2003 & 1628 & 1669 & 1781 & 2089 & 2434 \\
20 && 3098 & 2603 & 2614 & 2726 & 2966 & 3289 \\

\hline
\multicolumn{8}{c}{}\\
\multicolumn{8}{c}{(c) Quadratic approximation}\\
\hline
$\ell \backslash\epsilon$&& 0.1 &0.05& 0.01& 0.005& 0.001& 0.0005\\
\hline\hline
8 && 2078 & 2843 & 1125 & 1215 & & \\
12 && 4898 & 4928 & 2820 & 2895 & 2490 & 2265 \\
16 && 8925 & 8955 & 5940 & 6015 & 5280 & 5393 \\
20 && 11415 & 11438 & 10260 & 10335 & 9300 & 9405 \\
\hline
\end{tabular}
\end{center}
\end{table}

\subsubsection{Hybrid solution} The GC part takes care to detect the correct interval through the selection binary tree, retrieve the parameters, compute the difference between $\hat{x}$ and the left extreme of the segment and finally obfuscate the parameters and the difference. In total $3(2\sim(N-2)+\ell+\sum_{i=0}^{d}[(i+1)\ell_v+\tau])$ hash functions are evaluated by Bob and one quarter of them, on the average, by Alice.
After receiving the obfuscated values, Alice encrypts the powers of the difference ($d$ exponentiations), Bob removes the obfuscation ($\frac{d(d-1)}{2}$ exponentiations), then Alice performs $d-1$ decryptions ($d-1$ exponentiations), encrypts the parameters ($d+1$ exponentiations) and finally Bob removes the obfuscation affecting the approximation value with $2d$ exponentiations. 
In the linear implementation, the $\frac{d(d-1)}{2}$ exponentiations required to remove the obfuscation from the powers and the following $d-1$ decryptions are not needed.
\tab{tab:hybcomputation} shows the complexity of the hybrid protocol when linear and quadratic approximations are used. We can observe that many non-XOR gates are usually replaced by a fixed number of exponentiations. In some cases, especially with small $\ell$, the GC part of the hybrid protocol needs more Hashes than the corresponding full-GC solution, because obfuscation needs many bits. 
\begin{table*}[t]
\begin{center}
\caption{Average computational complexity of the online phase expressed as number of hashes ($H$) and exponentiations ($E$) (rounded to the closest integer) for both Alice and Bob in the hybrid approximation protocol applied to $\mathit{sinc}(x)$. Complexity is expressed as a function of $\ell$ and $\epsilon$ (missing values mean that the given error cannot be reached with the given bitlength).}
\label{tab:hybcomputation}
{
\begin{tabular}{ccrrrrrr}

\multicolumn{8}{c}{(a) Linear approximation}\\
\hline
$\ell \backslash\epsilon$&& 0.1 &0.05& 0.01& 0.005& 0.001& 0.0005\\
\hline\hline
8 && 1061$H$+5$E$ & 1106$H$+5$E$ & 1241$H$+5$E$ & 1339$H$+5$E$ & &  \\
12 && 1114$H$+5$E$ & 1166$H$+5$E$ & 1316$H$+5$E$ & 1429$H$+5$E$ & 1950$H$+5$E$ & 2625$H$+5$E$ \\
16 && 1174$H$+5$E$ & 1226$H$+5$E$ & 1376$H$+5$E$ & 1489$H$+5$E$ & 1980$H$+5$E$ & 2325$H$+5$E$ \\
20 && 1234$H$+5$E$ & 1286$H$+5$E$ & 1436$H$+5$E$ & 1549$H$+5$E$ & 2033$H$+5$E$ & 2355$H$+5$E$ \\

\hline
\multicolumn{8}{c}{}\\
\multicolumn{8}{c}{(b) Quadratic approximation}\\
\hline
$\ell \backslash\epsilon$&& 0.1 &0.05& 0.01& 0.005& 0.001& 0.0005\\
\hline\hline
8 && 1433$H$+11$E$ & 1504$H$+11$E$ & 1519$H$+11$E$ & 1609$H$+11$E$ & &  \\
12 && 1526$H$+11$E$ & 1568$H$+11$E$ & 1609$H$+11$E$ & 1684$H$+11$E$ & 1845$H$+11$E$ & 2111$H$+11$E$ \\
16 && 1631$H$+11$E$ & 1665$H$+11$E$ & 1699$H$+11$E$ & 1774$H$+11$E$ & 1905$H$+11$E$ & 2018$H$+11$E$ \\
20 && 1699$H$+11$E$ & 1725$H$+11$E$ & 1789$H$+11$E$ & 1864$H$+11$E$ & 1995$H$+11$E$ & 2100$H$+11$E$ \\
\hline
\end{tabular}
}
\end{center}
\end{table*}

Comparing the full-GC and hybrid solutions from a computational point of view is problematic, since this requires to compare the complexity of Hash functions and exponentiations, which ultimately depends on the architecture of the platform used to implement the protocols. In the next section, we move one step in this direction by comparing the runtime of two specific implementations of the protocols.

\subsection{Runtimes}  
We measured the runtimes required by Java implementations of the protocols on a desktop PC having an AMD Phenom II X4 p40 processor at 3.00 GHz and 6.00 GB of RAM.  The results are reported in \tab{tab:runtime}. 
According to such a table, GC solutions are preferable to Hybrid ones. However all the tests have been performed by running both the evaluator and the garbler in the same PC, so that the runtimes are more related to computational complexity rather than to communication time. \tab{tab:choiceRuntime} shows the preferable solution for various bitlength-precision setups.

\begin{table}[t]
\begin{center}
\caption{Average Garbler/Evaluator Online Runtimes (millisec) }
\label{tab:runtime}
\begin{tabular}{@{\hspace{0.2cm}}c@{\hspace{0.2cm}}c@{\hspace{0.2cm}}c@{\hspace{0.2cm}}c@{\hspace{0.2cm}}c@{\hspace{0.2cm}}c@{\hspace{0.2cm}}c@{\hspace{0.2cm}}c@{\hspace{0.2cm}}}

\multicolumn{8}{c}{(a) Full-GC Constant approximation}\\
\hline
$\ell \backslash\epsilon$&& 0.1 &0.05& 0.01& 0.005& 0.001& 0.0005\\
\hline\hline
8 && 0.6/0.3 & 1.0/0.4 & 2.6/1.2 & 3.6/1.7 & & \\ 
12 && 0.5/0.3 & 0.9/0.4 & 4.6/2.2 & 65/7.6 & 96/37 & 121/64 \\ 
16 && 0.5/0.2 & 1.0/0.4 & 4.8/2.2 & 87/10 & 131/57 & 180/107 \\ 
20 && 0.7/0.2 & 1.1/0.4 & 5.8/2.8 & 99/10 & 145/59 & 200/113 \\ 
\hline
\multicolumn{8}{c}{}\\
\multicolumn{8}{c}{(b) Full-GC Linear approximation}\\
\hline
$\ell \backslash\epsilon$&& 0.1 &0.05& 0.01& 0.005& 0.001& 0.0005\\
\hline\hline
8 && 2.4/1.1 & 1.9/0.8 & 2.0/0.9 & 2.4/1.0 & & \\ 
12 && 5.2/2.3 & 4.2/2.1 & 4.2/1.9 & 4.6/2.1 & 52/3.0 & 56/6.3 \\ 
16 && 72/6.3 & 67/4.5 & 70/4.3 & 70/4.7 & 72/5.9 & 74/8.3 \\ 
20 && 93/12 & 95/10 & 90/10.0 & 92/9.7 & 95/12 & 92/12 \\ 

\hline
\multicolumn{8}{c}{}\\
\multicolumn{8}{c}{(c) Full-GC Quadratic approximation}\\
\hline
$\ell \backslash\epsilon$&& 0.1 &0.05& 0.01& 0.005& 0.001& 0.0005\\
\hline\hline
8 && 31/2.9 & 30/2.9 & 3.4/1.5 & 3.6/1.4 & & \\ 
12 && 52/11 & 53/11 & 48/5.8 & 47/5.8 & 49/4.7 & 48/4.4 \\ 
16 && 81/23 & 82/23 & 77/16 & 76/15 & 71/13 & 75/14 \\ 
20 && 107/31 & 108/31 & 102/28 & 103/28 & 100/26 & 99/25 \\ 
\hline
\multicolumn{8}{c}{}\\
\multicolumn{8}{c}{(d) Hybrid Linear approximation}\\
\hline
$\ell \backslash\epsilon$&& 0.1 &0.05& 0.01& 0.005& 0.001& 0.0005\\
\hline\hline
8 && 486/310 & 479/306 & 500/323 & 496/322 & & \\ 
12 && 516/321 & 511/318 & 520/327 & 517/328 & 520/327 & 526/327 \\ 
16 && 536/327 & 541/327 & 536/324 & 533/321 & 542/329 & 541/330 \\ 
20 && 554/325 & 553/319 & 553/321 & 553/328 & 557/332 & 561/328 \\ 

\hline
\multicolumn{8}{c}{}\\
\multicolumn{8}{c}{(e) Hybrid Quadratic approximation}\\
\hline
$\ell \backslash\epsilon$&& 0.1 &0.05& 0.01& 0.005& 0.001& 0.0005\\
\hline\hline
8 && 1155/642 & 1161/644 & 1169/645 & 1163/644 & & \\ 
12 && 1187/643 & 1187/659 & 1195/655 & 1192/660 & 1182/648 & 1176/645 \\ 
16 && 1197/650 & 1189/644 & 1182/639 & 1194/645 & 1197/651 & 1198/653 \\ 
20 && 1226/660 & 1226/656 & 1227/660 & 1205/643 & 1219/650 & 1223/643 \\ 
\hline
\end{tabular}
\end{center}
\end{table}

\begin{table}[!h]
\begin{center}
\caption{Best protocol for each configuration (protocol - polynomial degree). }
\label{tab:choiceRuntime}
{
\begin{tabular}{ccrrrrrr}
\multicolumn{8}{c}{(a) Constant approximation}\\
\hline
$\ell \backslash\epsilon$&& 0.1 &0.05& 0.01& 0.005& 0.001& 0.0005\\
\hline\hline
 8 & &GC-0 &GC-0 &GC-1 &GC-1 &    &     \\
12 & &GC-0 &GC-0 &GC-1 &GC-1 &GC-2 &GC-2 \\
16 & &GC-0 &GC-0 &GC-0 &GC-1 &GC-1 &GC-1 \\
20 & &GC-0 &GC-0 &GC-0 &GC-1 &GC-1 &GC-1 \\
\hline
\end{tabular}
}
\end{center}
\end{table}

 \section{Conclusion}\label{sec:conclusion}
    Given a function $f()$ and an interval belonging to its domain, we considered the problem of approximating ${f}()$ by means of a piecewise polynomial function $\widetilde{f}()$ in a STPC setting. Constant, linear and quadratic approximations have been considered. Regardless of the polynomial degree (except for the constant approximation), two possible protocols have been proposed. The first one relies completely on Garbled Circuit theory, while the other adopyts a hybrid solution where GC and Homomorphic Encryption are used together.

The main advantage of the full-GC implementation is the use of only one cryptographic primitive. {If the function $f()$ is part of a protocol, where the previous and subsequent functionalities are also implemented by using GC's, the integration of the sub-protocols that approximate $f()$ would be very easy.}
The evaluation of the piecewise approximation is the heaviest part of the GC protocol and its complexity significantly increases with the degree of the polynomial.
The hybrid solution permits to evaluate the final part by using HE, reducing the complexity when polynomial with high degrees are used.

Communication and computational complexity have been estimated for a sample function. Runtimes have been measured as well. Observing runtimes, the full-GC solutions are preferable to Hybrid solutions and we highlight that generally a function approximation can be run in less than 100 msec. Anyway, from a communication point of view,  with high bitlengths and high precision, hybrid solutions are preferable to full-GC solutions, becoming more efficient in scenarios with low bandwidth or when the result is used following HE-based STPC protocols. Constant or linear approximation can be generally used, while we rarely observed an improvement from the use of quadratic approximation, making the use of higher polynomial degrees unpractical. 

Thanks to function approximation, many secure protocols can be optimized or even implemented for the first time. This is the case, for instance, of neural networks with smoother activation functions. Moreover a future extension to multivariate functions can be used to approximate an entire secure protocol performing a computation on several inputs made available by different parties.

\bibliographystyle{IEEEtran}
{\footnotesize
\bibliography{biblio/biblio}}


\end{document}